# Optical properties of thin-film vanadium dioxide from the visible to the far infrared

Chenghao Wan[1,2], Zhen Zhang[3], David Woolf[4], Colin M. Hessel[4], Jura Rensberg[5], Joel M. Hensley[4], Yuzhe Xiao[1], Alireza Shahsafi[1], Jad Salman[1], Steffen Richter[6,7], Yifei Sun[3], M. Mumtaz Qazilbash[8], Rüdiger Schmidt-Grund[6], Carsten Ronning[5], Shriram Ramanathan[3], Mikhail A. Kats[1,2]

[1] Department of Electrical and Computer Engineering, University of Wisconsin-Madison, Madison, Wisconsin 53706, USA
[2] Department of Materials Science & Engineering, University of Wisconsin-Madison, Madison, Wisconsin 53706, USA
[3] School of Materials Engineering, Purdue University, West Lafayette, IN 47907, USA
[4] Physical Sciences Inc., 20 New England Business Center, Andover, Massachusetts 01810-1077, USA
[5] Institute for Solid State Physics, Friedrich-Schiller-Universität Jena, 07743 Jena, Germany
[6] Universität Leipzig, Felix Bloch Institute for Solid State Physics, Linnéstr. 5, 04103 Leipzig, Germany
[7] ELI Beamlines/Fyzikální ústav AV ČR, v.v.i., Za Radnicí 835, Dolní Břežany, Czech Republic
[8] Department of Physics, College of William and Mary, Williamsburg, Virginia 23187-8795, USA

**Abstract**

The insulator-to-metal transition (IMT) in vanadium dioxide ($VO_2$) can enable a variety of optics applications, including switching and modulation, optical limiting, and tuning of optical resonators. Despite the widespread interest in optics, the optical properties of $VO_2$ across its IMT are scattered throughout the literature, and are not available in some wavelength regions. We characterized the complex refractive index of $VO_2$ thin films across the IMT for free-space wavelengths from 300 nm to 30 μm, using broadband spectroscopic ellipsometry, reflection spectroscopy, and the application of effective-medium theory.

We studied $VO_2$ thin films of different thickness, on two different substrates (silicon and sapphire), and grown using different synthesis methods (sputtering and sol gel). While there are differences in the optical properties of $VO_2$ synthesized under different conditions, they are relatively minor compared to the change resulting from the IMT, most notably in the ~2 – 11 μm range where the insulating phase of $VO_2$ has relatively low optical loss. We found that the macroscopic optical properties of $VO_2$ are much more robust to sample-to-sample variation compared to the electrical properties, making the refractive-index datasets from this article broadly useful for modeling and design of $VO_2$-based optical and optoelectronic components.

**Introduction**

Vanadium dioxide ($VO_2$) undergoes a first-order insulator-to-metal transition (IMT) at ~68 °C, which can be driven thermally [1], electrically [2], optically [3], or via strain [4]. This reversible phase transition is the result of an interplay between Mott- and Peierls-type mechanisms [5], and can result in carrier-density changes of up to four orders of magnitude across the IMT [6]. The dramatic change in optical properties of $VO_2$ that accompanies the carrier-density change [7] has advanced a variety of applications, including optical limiting [8][9], nonlinear isolation [10], switching [11][12], and thermal-emission engineering [13]. Depending on the $VO_2$ film quality, grain size, the concentration and type of potential impurities, and the degree of strain, the phase transition can vary from abrupt to gradual [14][15][16]. Either of these may be preferred depending on the application; e.g., abrupt transitions for switching [12], and gradual transitions for tuning of optical resonances [17].

Knowledge of the complex refractive index of $VO_2$ across its IMT can enable computational design of various optical and optoelectronic devices incorporating this material. There is a large literature base of experiments that have been used to extract such optical properties, most notably using spectroscopic ellipsometry [18]; however, many of these investigations focus almost entirely on the visible and near-infrared ranges [7][18][19][20][21], with one recent study extending to the mid infrared [22].

The goal of this paper is to provide a comprehensive study of the optical properties of thin-film $VO_2$ across its IMT for free-space wavelengths from 300 nm to 30 µm. We studied films of different thickness (from ~70 to ~130 nm), grown by different methods (magnetron sputtering and sol-gel synthesis), and on different substrates (silicon and sapphire). To extract the complex refractive-index data, we used variable-angle spectroscopic ellipsometry (VASE) [23], and verified the results with a combination of standard material-characterization techniques and non-ellipsometric reflectance measurements. To model the gradually changing optical properties throughout the IMT, we used a simple effective-medium theory, which explains most of our experimental observations. Taken together, the results of this article can be used to perform optical simulations of $VO_2$-containing devices over the broadband range covering the ultraviolet to the far infrared, for $VO_2$ in the insulating and metallic phases, as well as in the intermediate regime throughout the IMT.

**Methods**

We selected four $VO_2$ thin-film samples with varying thickness, substrate, and synthesis method (Table 1); the synthesis details are in *Supplementary Section 1*. These samples were chosen such that there were three pairs that primarily differ in only one way for each pair: Films 1 and 2 grown using the same technique and on the same substrate, but with substantially different thickness (highlighted in purple), Films 2 and 3 grown to a similar thickness but on different substrates (highlighted in green), and Films 2 and 4 grown to a similar thickness on the same substrate but using different synthesis techniques (highlighted in yellow). The thicknesses given in Table 1 were measured using cross-sectional scanning electron microscopy (SEM), and the uncertainties are surface-roughness values determined using atomic force microscopy (AFM).

**Table 1.** List of samples characterized in this work

|        | Thickness (nm) | Substrate          | Synthesis method       |
|--------|----------------|--------------------|------------------------|
| Film 1 | 70 ± 9         | Si + native oxide  | Magnetron sputtering   |
| Film 2 | 130 ± 17       | Si + native oxide  | Magnetron sputtering   |
| Film 3 | 120 ± 12       | Sapphire           | Magnetron sputtering   |
| Film 4 | 110 ± 5        | Si + native oxide  | Sol-gel                |

The thickness of the $VO_2$ film may affect its optical properties due to strain relaxation (the strain-relaxation thickness is in the tens of nanometers [24][25]). To investigate this difference, we prepared $VO_2$ films on silicon (001) substrates (with a native oxide layer) with thickness of ~70 and ~130 nm (Films 1 and 2) using the same magnetron-sputtering recipe.

Commonly, both silicon (001) and *c*-plane-oriented sapphire are used as substrates for $VO_2$ growth [26]. The substrate choice can significantly affect the film quality, in part due to the lattice mismatch at the substrate-film interface [15][26]. To explore the role of the substrate on the optical properties, we prepared sputtered films of similar thickness (~120 and ~130 nm), but on different substrates: silicon (001) with a native oxide layer (Film 2), and *c*-plane-oriented sapphire (Film 3). Film 3 had lower roughness (*Supplementary Section 2*) and a larger change in resistivity across the IMT (*Supplementary Section 3*), which is consistent with the lower lattice mismatch on sapphire.

Finally, the growth technique can also have a large influence on the properties of $VO_2$ [27]. Many synthesis techniques have been used for $VO_2$ growth, including sputtering [26], sol-gel synthesis [28], atomic-layer deposition [29], chemical-vapor deposition [30], and pulsed-laser deposition [31]. Here, we compare films grown to a similar thickness (~110 and ~130 nm) on the same substrate (silicon with a native oxide layer), using magnetron-sputtering (Film 2) and the sol-gel method (Film 4).

To extract the complex refractive indices of the films, we performed two sets of VASE measurements, the first at shorter wavelengths (300 nm to 2 µm, using the V-VASE instrument from J. A. Woollam Co.), and the second at longer wavelengths (2 µm to 30 µm, IR-VASE). Both sets of measurements were performed at three incident angles of 50°, 60°, and 70°. Prior to the measurements, the back surfaces of the substrates were sandblasted to minimize backside reflections.

The ellipsometry measurements were carried out only at two temperatures: 30 °C ($VO_2$ in the insulating phase) and 100 °C (in the metallic phase). For intermediate temperatures, we combined the ellipsometry data with temperature-dependent reflection measurements, and calculated the temperature-dependent effective refractive index using an effective-medium theory (see next section). We chose this approach because ellipsometry across the IMT can result in artifacts due to temperature fluctuations of the sample combined with the hysteretic nature of the IMT; these artifacts are worse for ellipsometry than the reflectivity measurements because of the longer duration of a single ellipsometry measurements at a given temperature. The reflectance measurements were performed at near-normal incidence, for wavelengths from 2 µm to 17 µm, using an infrared microscope (Bruker Hyperion 2000) attached to a Fourier-transform infrared spectrometer (Bruker Vertex 70). All reflectance spectra were collected at temperatures between 30 °C and 100 °C (first heating, then cooling) with steps of 2 °C, and were normalized to a gold mirror, assuming a known reflectance of optically thick gold.

**Data analysis and discussion**

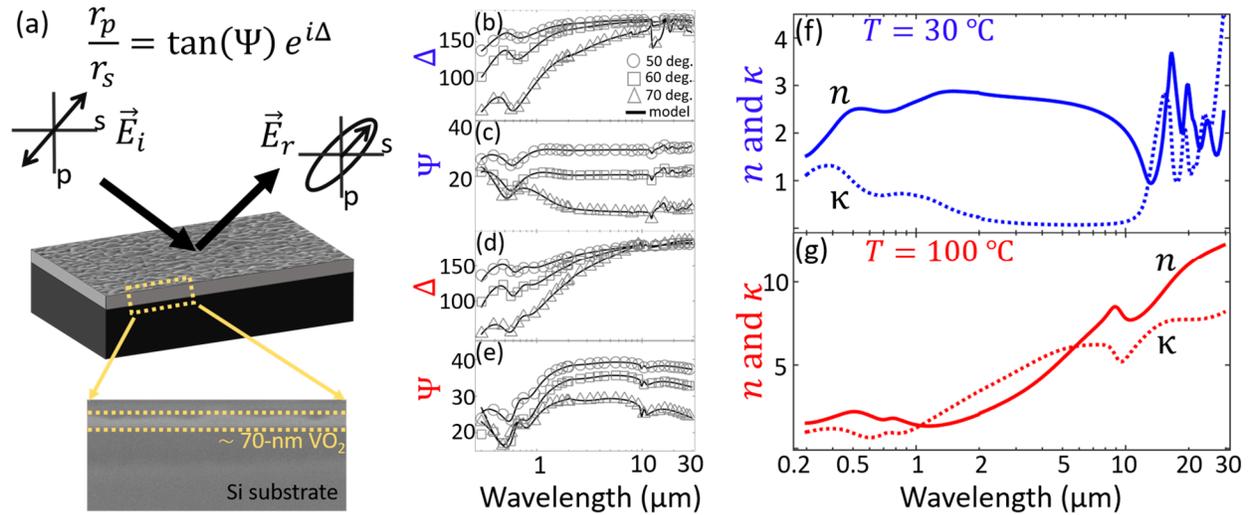

**Figure 1.** (a) Schematic of our ellipsometry measurement. The inset is an SEM cross section of Film 1, including the definitions of the ellipsometric measured data, Ψ and Δ. Figures (b – e) are the experimental (discrete points) and fitted (solid curves) Ψ and Δ for the (b, c) insulating (30 °C) and (d, e) metallic phase (100 °C) of Film 1. (f, g) Extracted real ($n$) and imaginary ($\kappa$) parts of the complex refractive indices of Film 1 in its insulating and metallic phase, respectively, where our fitting process assumed no surface roughness, silicon-oxide layer, or potential non-stoichiometric contributions from other vanadium oxides.

Our fitting procedure comprised the following basic steps. First, we fit the shorter-wavelength V-VASE data, obtaining the complex refractive index and the fitted thickness of the $VO_2$ film. The shorter-wavelength fit was carried out before moving to longer wavelengths because, for deeply subwavelength films, there may be no unique fit for both the thickness and complex refractive index [32][33]. Therefore, it is advantageous to first fit at only the shorter wavelengths. Then, we used the thickness from the V-VASE fit as an input parameter into the fitting model for the longer-wavelength IR-VASE data. Finally, we confirmed that the resulting complex refractive index matched well between the two datasets in the spectral range where the two overlap, and that the fitted thickness matched the SEM/AFM measurements. We performed complete analysis of all of the films listed in Table 1, but for brevity only present all of the ellipsometric data for Film 1 in the main text (see *Supplementary Section 4* for the remaining data).

The experimental ellipsometric data ($\Psi$ and $\Delta$ in Fig. 1a) from the V-VASE are shown in Fig. 1(d-g) for Film 1 in both the insulating (30 °C) and metallic (100 °C) phase. Our fitting model consisted of a semi-infinite silicon substrate and the $VO_2$ film. The optical constants of the silicon substrate were taken from Ref. [34], and the native oxide layer (thickness ~3 nm) of silicon was neglected in the fitting, because it had no noticeable impact on the resulting fits (*Supplementary Section 4*). The thickness of the $VO_2$ layers was a fitting parameter for the insulating-phase $VO_2$, and we assumed that the thickness change during the transition (expected to be ~0.6% [35][36]) was much smaller than our thickness uncertainty. We parametrized the model dielectric function for the $VO_2$ layer using a series of Lorentzian oscillators [37][38] for the insulating phase, and additional Drude term to capture the contribution of the free carriers [37][39] for the metallic phase. The parameters of these oscillator terms were fitted. We ignored the presence of surface roughness (arithmetic average roughness $R_a$ ~ 9 nm based on AFM imaging), which is much smaller than the measurement wavelength. For analysis that explicitly considers the surface roughness, see *Supplementary Section 4*.

The V-VASE fitted thickness of $VO_2$ (~69 nm) agrees with the cross-sectional SEM imaging to within the surface roughness (~9 nm, by AFM, see Supplementary Section 3). This fitted thickness enabled us to initialize our IR-VASE modeling with a fixed effective optical thickness and no surface roughness layer. The resulting fitting of the IR-VASE data enabled us to obtain the real ($n$) and imaginary ($\kappa$) parts of the complex refractive index throughout the mid- and far-infrared regions [Fig. 1(f, g)].

Combining the results of the V-VASE and IR-VASE data analysis, we plotted the complex refractive indices of Film 1 in both the insulating and metallic phases [Fig. 1(f, g)] at 30 °C and 100 °C, respectively. The transition between the spectral regions of the two ellipsometers ($\lambda$ ~ 2 µm) is hardly noticeable in the data, indicating good agreement between these two separately fitted results. To further confirm the extracted complex refractive indices, we calculated the normal-incidence reflectance of this sample using the transfer-matrix method, which agrees well with FTIR reflectance measurements for both $VO_2$ phases (see Fig. 4 and the corresponding text below).

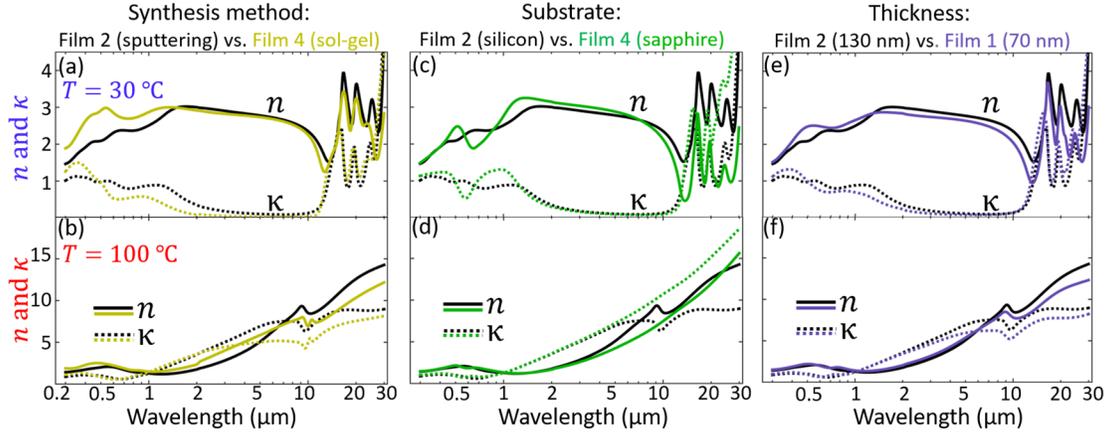

**Figure 2.** Comparison of the extracted refractive indices of the VO$_2$ films in Table 1 in the insulating (top) and metallic (bottom) phase, by different (a, b) synthesis methods; (c, d) substrates; and (e, f) film thicknesses. Note that the comparisons we make here are between the four films in Table 1, and the results should not be interpreted as definitive for, e.g., differences between all sputtered and sol-gel-synthesized VO$_2$ films.

We applied this characterization procedure to the other three samples, to compare the optical properties of VO$_2$ films for different synthesis methods [Fig. 2(a, b)], substrates [Fig. 2(c, d)], and thicknesses [Fig. 2(e, f)]. We note that for Film 3, grown on *c*-plane-oriented sapphire, our fitting procedure differs from the one used for all of the other films due to the anisotropy and significant dispersion of sapphire in the mid- and far infrared. Details of the modified fitting procedure are discussed in *Supplementary Section 4*.

It is instructive to examine the differences and similarities between the optical properties of these four films, focusing on different wavelength ranges. For wavelengths below ~2 µm, which approximately corresponds to the energy region of interband transitions of VO$_2$ in its insulating phase (>0.6 eV [40]), all of our films show substantial differences in their insulating-phase optical properties. Such large differences of optical properties in the visible and near-infrared ranges are also regularly found in the literature for different VO$_2$ samples [18][19][22][41][42][43][44] (Fig. 3). The band structure of the insulating state can be sensitive to a variety of factors, including strain, stoichiometry, grain size, and defects [15][26]. All of these factors change with varying growth conditions [45], lattice matching to the substrate [15], and film thickness (and the resulting strain relaxation) [25][46]. The short-wavelength differences may also be in part due to the significant differences in the surface roughness, as seen by our SEM and AFM analysis (*Supplementary Section 2*). As discussed in *Supplementary Section 4*, there were slight changes in $n$ and $\kappa$ when we considered surface roughness in our ellipsometry modeling.

In the long-wavelength region accessible to our measurements (11-30 µm), insulating-phase VO$_2$ films feature three strong vibrational resonances (at approximately λ = 17, 20, and 25 µm [47]). These long-wavelength features do not appear to vary very much for different synthesis techniques [Fig. 2(a)], do vary somewhat for different thickness [Fig. 2(e)], and are quite distinct for different substrates [Fig. 2(c)], probably due to different strain resulting from the lattice mismatch with the substrate. VO$_2$ crystal orientation, which varies significantly for films on different substrates, can also affect such phonon features [48]. In the metallic phase, vibrational resonances are not observed, as the Drude contribution dominates. We are unsure about the origin of the feature around 10 µm in the metallic phase, though the wavelength corresponds to a vibrational resonance in V$_2$O$_5$ [49] which may be present in some of our films. It is not clear why the feature disappears from our data for the insulating phase of VO$_2$, however it is possible that a surface layer of V$_2$O$_5$ is present at high temperature, but undergoes a reversible phase transition to another stoichiometry as the temperature decreases [50]. We also note that similar features can be found in some previously published experimental results, including for $VO_2$ films on sapphire, silicon, silica, and zinc oxide (ZnO) [17][51]. This feature is quite apparent on Si substrates in our data, but is much less noticeable for the film on sapphire [Fig. 2(d)].

Despite these differences, the overall trend of the refractive index for both insulating and metallic phases are quite consistent between all of our films. In particular, the refractive-index values are very similar between all of the films in the insulating phase in the mid-infrared region from 2 to 11 µm, which is also the region of the lowest optical losses. This is in stark contrast to electrical measurements using macroscopically spaced electrodes (see details in *Supplementary Section 2, 3*), which result in dramatic differences between the films: Film 1 shows a resistance change of ~2 orders of magnitude across the IMT; Film 2 shows ~3 orders; Film 3 shows ~4 orders; and in Film 4, no change could be observed at all, likely due to the small-scale cracks in the film (see *Supplementary Section 2*). Such structural features impede macroscopic electrical conduction, but are much smaller than the scale of the optical wavelength (> 300 nm), and therefore their impact on the optical properties is expected to be relatively minor.

For a more-comprehensive comparison, we plot the complex refractive index of Film 1, together with data from a handful of literature references [18][19][22][41][42][43][44], in Fig. 3. Most existing characterization works focus on the visible and near-infrared region, in which the optical properties of $VO_2$ differ substantially, as discussed previously. In the mid-infrared region, there are much fewer such measurements. The data from ref. [41] (from 1996) differs substantially from our results, and does not seem consistent with other literature data from $\lambda \sim$ 2 to 7 µm. Our data agrees well with ref. [42], though in that work the fitting process may not have fully accounted for the long-wavelength vibrational resonances of $VO_2$. Our data also agrees reasonably with the very recent ref. [22], which covers wavelengths up to 15 µm, though the metal-phase $VO_2$ from this work appears to be less metallic compared to our films.

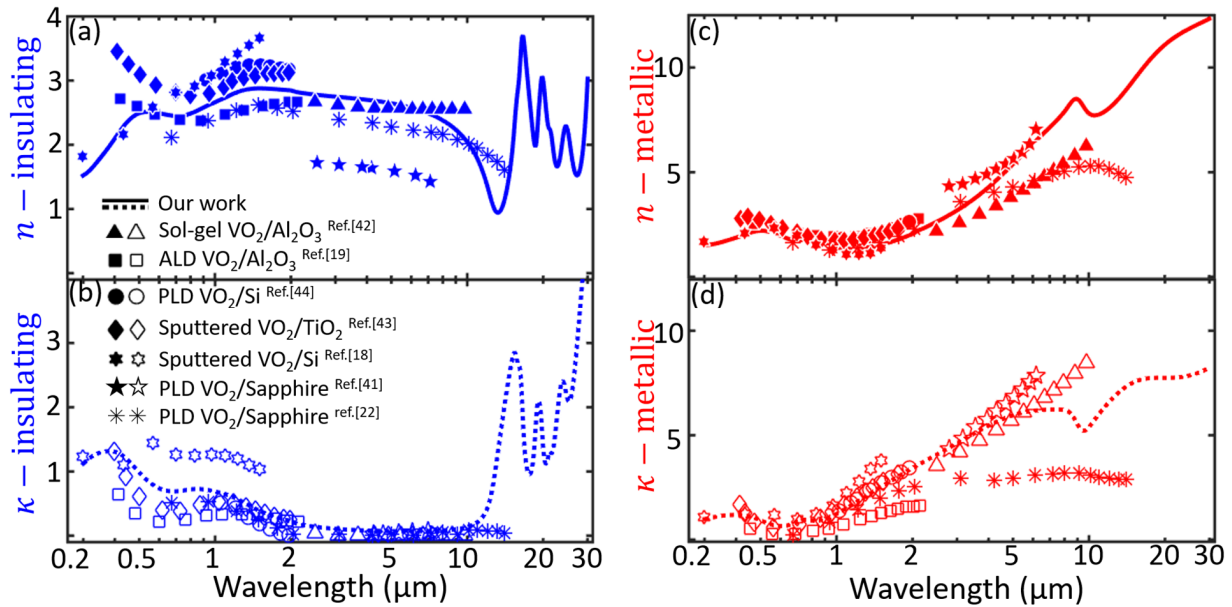

**Figure 3.** Comparison between the measured complex refractive index of Film 1 and some representative published data: (a, b) real and imaginary parts of the refractive index of insulator-phase thin-film $VO_2$; (c, d) real and imaginary part of the metal-phase $VO_2$ films. The deposition technique and substrate for each literature dataset are given in the legend. Thickness of all films is on the order of 100 nm.

In Figs. 2 and 3, we presented refractive-index data for the insulating and metallic phases. In principle, similar measurements can be performed for any temperature throughout the IMT. However, this can be challenging because these measurements require very stable temperature control due to the hysteretic nature of the IMT and the relatively slow ellipsometric measurements. Nevertheless, intermediate-state refractive-index data is often needed for the modeling of optical and optoelectronic devices that use the gradually tunable optical properties of $VO_2$ (e.g., [17]).

It has been demonstrated that effective-medium theories (EMTs) such as the Bruggeman formalism [52] and the Looyenga mixing rule [16][53] can be used to approximate the refractive indices of $VO_2$ films within the IMT, when insulating and metallic domains coexist. Here, we use the simpler Looyenga rule:

$$\tilde{\varepsilon}_{eff}{}^s = (1-f)\tilde{\varepsilon}_i{}^s + f\tilde{\varepsilon}_m{}^s \quad (1)$$

where $\tilde{\varepsilon} = \tilde{n}^2 = (n+i\kappa)^2$ is the complex dielectric function of $VO_2$, $f$ is the temperature-dependent volume fraction of the metal-phase $VO_2$ domains within the film, and $s$ varies from -1 to 1 depending on the shape of the metallic inclusions. We used the empirical value of $s = 0.3$ for thin-film $VO_2$ [16]. The effective refractive index is plotted in Fig. 4(a) for different values of $f$, given the complex-refractive-index data for Film 1.

The phase co-existence can be understood as a first-order equilibrium, and therefore the temperature dependence of $f$ can be expressed as [16][52]:

$$f(T) = \frac{1}{1+\exp\left[W/k_B\left(1/T - 1/T_{half}\right)\right]} \quad (2),$$

where $W$ contains information about the width of temperature range of the IMT, and $T_{half}$ is the temperature at which half of the volume of the film is in the metallic state. Note that due to the hysteresis in $VO_2$, the value of $T_{half}$ is different for heating and cooling (in fact, there may be an infinite number of values of $T_{half}$, if minor loops within the hysteretic region of $VO_2$ are considered [54]).

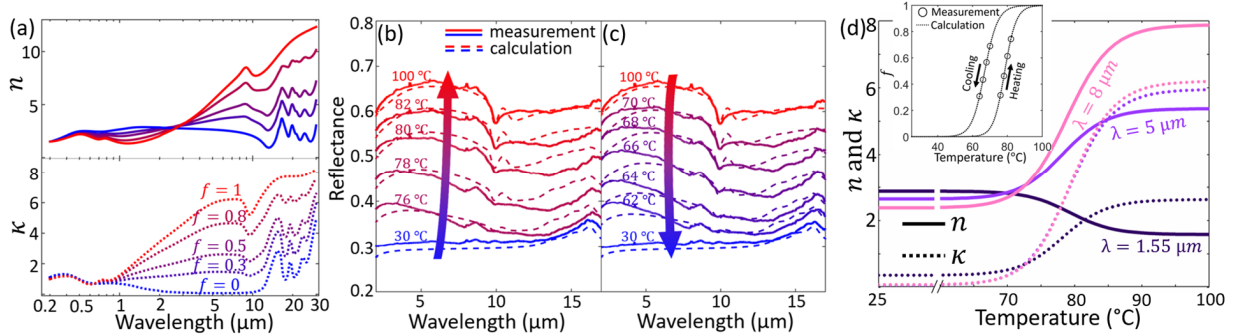

**Figure 4.** (a) The complex refractive index of $VO_2$ throughout its IMT, estimated using the Looyenga effective-medium theory (EMT) and the ellipsometry fits for Film 1. The values were calculated for volume fractions of the metallic phase $f$ = 0, 0.3, 0.5, 0.8, and 1, which correspond to increasing temperatures. Experimental (solid lines) and calculated (dashed lines) reflectance of Film 1 during (b) heating and (c) cooling, with a ramping rate of 1 °C/min. (d) Temperature-dependent refractive indices of Film 1 at λ = 1.55, 5, and 8 μm. The inset is the extracted temperature-dependent $f$ for a cycle of heating and cooling for Film 1.

For a given $W$ and $T_{half}$, Eqns. (1, 2) can be used to calculate the temperature-dependent $\tilde{n}(T)$, which can then be used for any conventional optics calculations (e.g., Fresnel equations and the transfer-matrix formalism). To determine $W$ and $T_{half}$ during the heating and cooling stages, we fit a calculation of the normal-incidence temperature-dependent reflectance in the mid infrared to our FTIR measurements, taken at temperature steps of 2 °C. The best fit values for Film 1 – 4 are listed in Table 2 and we chose Film 1 as a representative to show the agreement with the experiments [Fig. 4(b, c); for Film 2 – 4, see *Supplementary Section 5*].

**Table 2**. List of $W$ and $T_{half}$ values:

|  | Film 1 | Film 2 | Film 3 | Film 4 |
|---|---|---|---|---|
| $W$ (heating) | 3.37 eV | 3.57 eV | 4.85 eV | 9.60 eV |
| $W$ (cooling) | 2.75 eV | 3.79 eV | 4.15 eV | 7.14 eV |
| $T_{half}$ (heating) | 78.5 °C | 78.1 °C | 75.1 °C | 72.1 °C |
| $T_{half}$ (cooling) | 67.1 °C | 66.0 °C | 72.2 °C | 65.7 °C |

Once $f(T)$ is determined (e.g., the inset of Fig. 4d), one can obtain the complex refractive index at any wavelength of interest across the IMT, as shown in Fig. 4(d) where we picked three free-space wavelengths (1.55, 5, and 8 µm) to show the evolution of $n$ and $\kappa$ with increasing temperature.

**Conclusions**

By using a combination of variable-angle spectroscopic ellipsometry (VASE) and effective-medium theory (EMT), we extracted temperature-dependent complex-refractive-index values for vanadium-dioxide (VO₂) thin films over the wavelength range from 300 nm to 30 µm. We compared the results for VO₂ films of different thickness, on different substrates, and grown using different synthesis methods. We found that there were very large differences in electrical properties among the films, substantial differences in the optical properties at wavelengths below 2 µm, but relative consistency in the mid and far infrared, especially in the 2 – 11 µm region, which also corresponds to low optical losses for the insulating phase. Our full datasets, provided in the *Supplementary Information*, will be useful for those seeking to perform simulations of optical and optoelectronic devices based on VO₂.

**Acknowledgements**

We acknowledge support from the Office of Naval Research (MK: N00014-16-1-2556, SR: N00014-16-1-2398), the National Science Foundation (MMQ: DMR-1255156), and the Deutsche Forschungsgemeinschaft (CR: RO1198/21-1). Some of the fabrication and experiments were performed at the Materials Science Center (MSC) and the Soft Materials Laboratory (SML), both shared facilities managed by College of Engineering at the University of Wisconsin-Madison.

Supplementary Information:

Optical properties of thin-film vanadium dioxide from the visible to the far infrared

## Section 1. Synthesis of VO$_2$ films

Magnetron sputtering

Films 1, 2, and 3 were deposited onto single-side-polished *c*-plane sapphire and undoped (resistivity > 1000 $\Omega \cdot$ cm) Si (001) wafers via magnetron sputtering from a V$_2$O$_5$ target, with radio-frequency power of 100 W. During deposition, the chamber pressure was maintained at 5 mTorr with an Ar/O$_2$ gas mixture at a flow rate of 49.85/0.15 sccm. The substrate was heated to 700 °C for the formation of the VO$_2$ phase. The thicknesses of the resulting films are given in Table 1.

Sol-gel method

Our process generally follows the 'aqueous sol-gel' route established by Hanlon et al. [S1]. The method involves synthesizing the sol precursor by adding molten V$_2$O$_5$ to water. To produce the precursor sol, a crucible containing V$_2$O$_5$ powder is heated to 800 °C until the material melts. The molten V$_2$O$_5$ is then poured into water. The solution cools to room temperature and is filtered to remove the remaining solids. The solution is then spun onto the silicon substrate. Spin-coating is performed at between 1000 and 3500 rpm, depending on the viscosity of the sol and the desired film thickness. The substrate is baked for 5 mins at 200 °C to remove excess water. The samples are then thermally annealed at 550 °C in a tube furnace for 30 minutes. A mixture of 5%H$_2$/95%Ar forming gas is used as the flow gas.

## Section 2. SEM and AFM characterization

We imaged our samples with a scanning electron microscope (SEM, Zeiss LEO 1530) to confirm the continuity and uniformity of the films [Fig. S1(a–d)]. We also used a focused ion beam (FIB) to mill through the VO$_2$ films, so that we could image the cross sections [Fig. S1(a–d)]. Atomic force microscopy (AFM, Veeco MultiMode SPM) was used to extract the surface roughness of each sample [Fig. S1(a–d)].

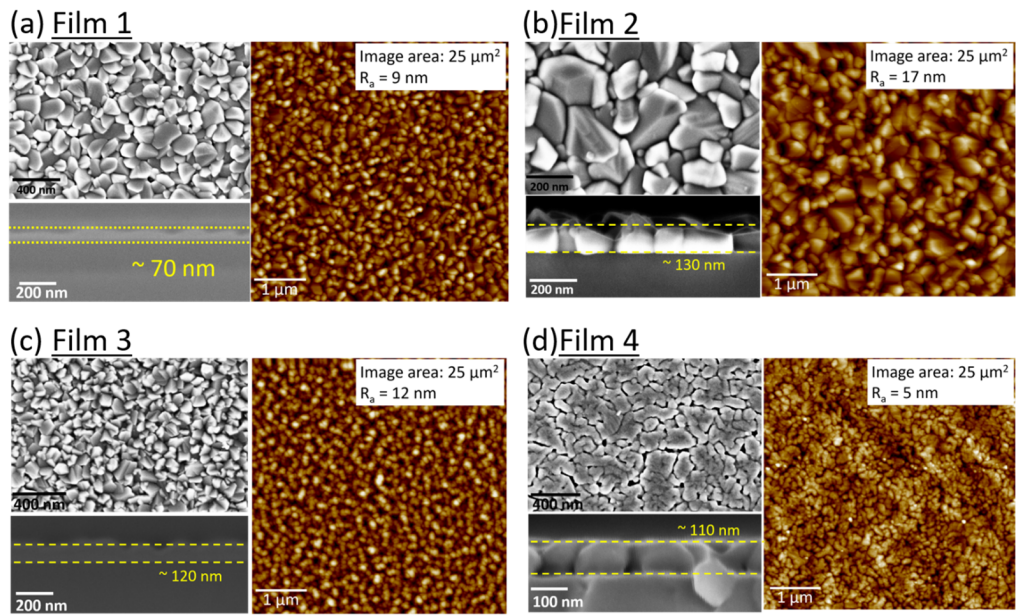

**Figure S1.** (a – d) SEM and AFM characterization of Films 1, 2, 3 and 4, respectively. In each of figures, the top left is the SEM image from the top; the bottom left is the SEM image of the cross section; the right is the AFM image, in which the surface roughness is listed in the top-right corner.

**Section 3. Resistivity measurements**

Post-growth temperature-dependent electrical resistance measurements were performed on the sputtered films on a customized temperature-controlled probe station. The resistance was obtained by measuring current while sweeping the voltage from -0.1 V to 0.1 V, using a Keithley 2635A source meter.

Film 3 features the largest change in resistivity across the IMT (Fig. S2), indicating that it has the best film quality, likely due to the smaller lattice mismatch at the interface between $VO_2$ and *c*-plane-oriented sapphire. For Film 4, we were not able to see the change in resistance across the IMT between two ~5-mm-separated electrodes, and the measured resistance is higher than that of the other films by several orders of magnitudes, so the data for this film is not shown. This large resistance in Film 4 is likely due to the discontinuities (small-scale cracks) in the film [Fig. S1(d)]. In general, such significant differences in electrical properties are likely due to variations in the defects, grain sizes, continuity at grain boundaries, surface roughness, etc., each of which may partially influence the macroscopic current path in the films, since distance between the electrodes is on the order of several millimeters. Despite such significant differences in electrical properties, all of the four films featured similar changes in their optical properties (e.g., reflectance shown in Fig. S6) throughout the IMT, because these microscopic structural differences are on a scale much smaller than the wavelength.

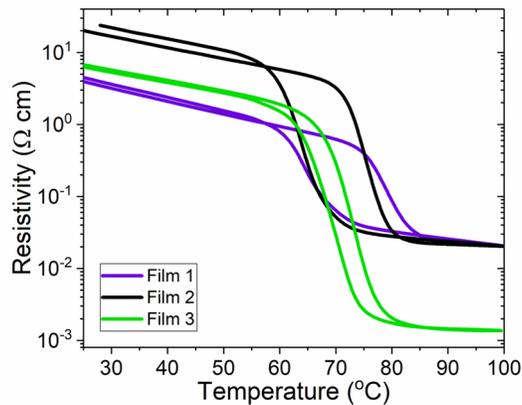

**Figure S2**. Temperature-dependent measurements of resistivity across the IMT for Film 1 – 3. Samples were thermally driven under a cycle of heating and cooling.

**Section 4. Ellipsometry modeling**

The parameterization that we used to model the dielectric function of $VO_2$ in its insulating phase include several Lorentz oscillators [S2] in the visible and near-infrared regions. Psemi-M0 functions, which are distributions of oscillators that can be used to model complicated asymmetric dielectric functions [S3][S4][S5], were used for the mid- and far-infrared wavelengths. For the metallic phase, an additional Drude term [S6] was added to describe the significantly increased carrier concentrations. Note that the choice of line shapes that results in the desired calculated Ψ and Δ is not unique for a particular experimental dataset [S2]. One can, in principle, use other line-shape functions to obtain equally good fitting results. However, the resulted $n$ and $\kappa$ values should be identical.

In Fig. S3(a), we show a fitting result for Film 1 assuming the presence of a surface roughness layer (50% air + 50% of the material underneath, by volume). We compare the fitting results between models with- and without the surface-roughness layer. The model without surface roughness is the same as the one described in the main text [Fig. 1(f) and black lines in Fig. S3(a)]. We added a surface roughness layer with a fixed thickness of 9 nm, as shown in Fig. S1(a), and used the same fitting procedure to obtain the refractive indices [purple lines in Fig. S3(a)]. The differences introduced by the surface roughness are more significant in the visible region and become ignorable for wavelengths beyond ~1.5 µm.

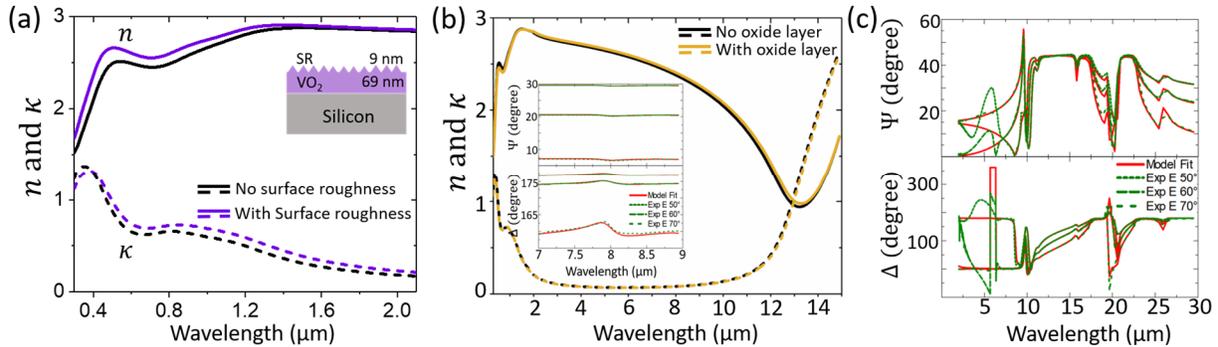

**Figure S3.** (a) Comparison of fitting results between two-layer (without a surface-roughness layer) and three-layer (with surface roughness) models. The fitted $n$ and $\kappa$ from the two-layer model (black lines) are the same as that in Fig. 1(f) in the main text. The fitted $n$ and $\kappa$ from the three-layer model that includes a 9-nm roughness layer are shown in purple. (b) Comparison of fitting results between modeling with and without the silicon-oxide layer. The fitted $n$ and $\kappa$ without $SiO_2$ layer (black lines) are the same as that in Fig. 1(f) in the main text. The fitted $n$ and $\kappa$ with a 3-nm $SiO_2$ layer (yellow lines) are essentially identical to the results from the modeling without a $SiO_2$ layer. The inset shows the $\Delta$ and $\Psi$ around 8 µm, where $SiO_2$ has a strong resonance, verifying the ~3-nm thickness of the $SiO_2$ layer. (c) Measured and fitted $\Psi$ and $\Delta$ of a bare $c$-plane-oriented sapphire wafer. The mismatches from 2 to 6.5 µm are due to the unintended reflection from the backside of sapphire substrate, which is transparent or semi-transparent in this wavelength range.

For films on silicon substrates, we need to consider the native oxide layer ($SiO_2$) between the $VO_2$ and silicon. As shown in the inset of Fig. S3(b), we added the oxide layer in our model, set its thickness as a fitting parameter and fit the ellipsometric data at wavelengths from 7 to 9 µm, since within this region $SiO_2$ features strong vibrational resonances [S7] so that its thickness could be well determined. We found the oxide layer is approximately 3 nm in all of our samples, and has no noticeable influence on the fitted values of $n$ and $\kappa$ of the $VO_2$ films [Fig. S3(b)]. Note that our cross-sectional SEM images [Fig. S1] also indicate that there are no thick silicon-oxide layers on any samples with a silicon substrate. Therefore, we chose to ignore the oxide layer in our final model, as discussed in the main text.

For the analysis of $VO_2$ films on sapphire, we have to model the substrate as uniaxial [S5]. As shown in Fig. S3(c), our modeling resulted in good fits of the $\Psi$ and $\Delta$ of a bare sapphire wafer, except in the ~2-7 µm wavelength range. In this range, sapphire is transparent or semi-transparent, resulting in poor fits due to light reflected from the back surface of the wafer (even though the wafer is single-side polished, long-wavelength light is insufficiently scattered). We attempted to remedy this via sandblasting of the back surface; however, we were not able to achieve a sufficient degree of roughness because the sapphire has a much higher hardness than silicon, and we did not want to risk damaging the sample by sandblasting at a high flow rate.

With the $VO_2$ layer on top (e.g., Film 3), the fitting procedure was different than the others due to the optical complexity (i.e., significant dispersion due to optical phonons, as well as anisotropy) of sapphire in the mid- and far-infrared regions. The fitting procedure used for silicon substrates and described in the main text is likely to artificially imprint the spectral features of sapphire into the multi-oscillator function of the $VO_2$ layer. We also found that it was difficult to find a good fit if the initial fitting parameters were not set close enough to a realistic solution. However, if we chose an initial multi-oscillator function that was the result of the fitting of another sample (e.g., Film 1), our fitting procedure converged.

Due to reflections from the back surface of sapphire, we discarded the data from 2 to 7 µm for the insulating phase of Film 3. Instead, we simultaneously fit the parameters VO$_2$ layer to the 7-30 µm region and the 1.6-2.1 µm region, because thin-film VO$_2$ is not expected to have any sharp features in the excluded region (e.g., Fig. 2). To verify the validity of our approach, we compared normal-incidence reflection measurements that included this region to calculations that use the fitted data [Fig. S5(c)] and observed excellent agreement. Metal-phase VO$_2$ is sufficiently lossy that the back-surface reflections could be neglected at all wavelengths.

The resulting complex refractive index for all films is shown in Fig. 2 in the main text, with all four sets of curves reproduced for clarity in Fig. S4. Tables with the data used to make these plots can be found in *Section 6*.

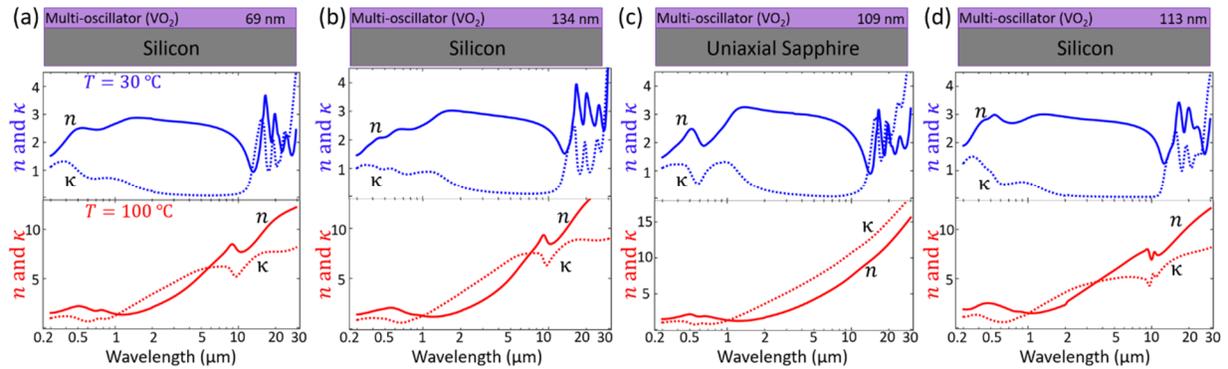

**Figure S4**. (a – d) ellipsometry-fitted refractive indices of Films 1, 2, 3, and 4 in both insulating and metallic phases.

### Section 5. Effective-medium calculations

As described in the main text, we used the Looyenga effective-medium theory to calculate the complex refractive index of VO$_2$ throughout its IMT. The resulting data for all four films are shown in Fig. S5; Fig. S5(a) is the same as Fig. 4(a, b) in the main text. Note that due to the hysteresis in VO$_2$, we characterized the evolution of refractive indices for both heating and cooling processes, as discussed in the main text. Here we only show the comparison between the calculations and measurements during heating. The fitting parameters ($W$ and $T_{half}$) extracted using the FTIR reflectance spectra are shown in Table 2 in the main text, for both heating and cooling.

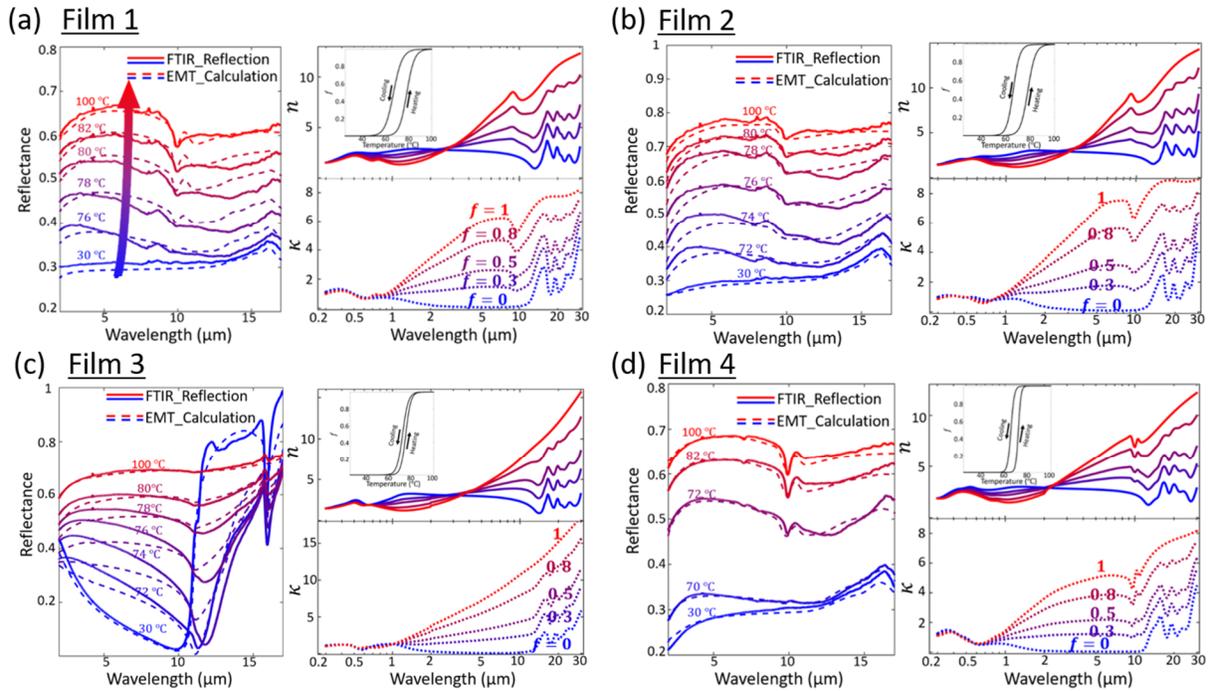

**Figure S5**. Left figures in (a – d) are comparisons of EMT-fitted and measured reflectance. Right figures in (a – d) are EMT-calculated $n$ and $\kappa$ for different metal fractions ($f$) throughout the IMT.

## Section 6. Tables of complex refractive indices of Films 1 – 4 in insulating and metallic phases

Note: the VO$_2$ complex-refractive-index data here was obtained from the fitting procedure described in the main text, and corresponds to Figs. 1(f, g), 2(a-f), and the solid curve in Fig. 3. As described above and in the main text, the fits resulting in this data assumed no surface roughness, nature silicon-oxide layer, or potential non-stoichiometric contributions from other vanadium oxides (e.g., V$_2$O$_5$).

| WAVELENGTH (MICRON) | FILM 1 $n$ (insulating) | FILM 1 $\kappa$ (insulating) | FILM 1 $n$ (metallic) | FILM 1 $\kappa$ (metallic) | FILM 2 $n$ (insulating) | FILM 2 $\kappa$ (insulating) | FILM 2 $n$ (metallic) | FILM 2 $\kappa$ (metallic) |
|---|---|---|---|---|---|---|---|---|
| 0.3 | 1.52 | 1.12 | 1.52 | 0.99 | 1.47 | 1.01 | 1.43 | 0.88 |
| 0.31 | 1.56 | 1.17 | 1.53 | 1.02 | 1.49 | 1.05 | 1.44 | 0.92 |
| 0.32 | 1.61 | 1.22 | 1.55 | 1.06 | 1.53 | 1.08 | 1.46 | 0.95 |
| 0.33 | 1.66 | 1.25 | 1.58 | 1.09 | 1.59 | 1.11 | 1.49 | 0.98 |
| 0.34 | 1.72 | 1.28 | 1.62 | 1.12 | 1.65 | 1.12 | 1.53 | 1.01 |
| 0.35 | 1.79 | 1.29 | 1.66 | 1.14 | 1.70 | 1.13 | 1.56 | 1.03 |
| 0.36 | 1.85 | 1.31 | 1.70 | 1.16 | 1.76 | 1.12 | 1.59 | 1.04 |
| 0.37 | 1.91 | 1.31 | 1.74 | 1.17 | 1.80 | 1.11 | 1.63 | 1.06 |
| 0.38 | 1.97 | 1.32 | 1.78 | 1.18 | 1.84 | 1.11 | 1.66 | 1.07 |
| 0.39 | 2.03 | 1.31 | 1.83 | 1.18 | 1.88 | 1.10 | 1.70 | 1.08 |
| 0.4 | 2.09 | 1.31 | 1.87 | 1.18 | 1.92 | 1.10 | 1.74 | 1.08 |
| 0.41 | 2.15 | 1.29 | 1.92 | 1.18 | 1.96 | 1.09 | 1.78 | 1.07 |
| 0.42 | 2.21 | 1.27 | 1.96 | 1.17 | 2.00 | 1.07 | 1.81 | 1.07 |
| 0.43 | 2.27 | 1.25 | 2.00 | 1.16 | 2.03 | 1.05 | 1.84 | 1.06 |
| 0.44 | 2.32 | 1.21 | 2.05 | 1.14 | 2.05 | 1.03 | 1.86 | 1.05 |
| 0.45 | 2.36 | 1.18 | 2.09 | 1.12 | 2.07 | 1.00 | 1.88 | 1.04 |
| 0.46 | 2.40 | 1.14 | 2.12 | 1.09 | 2.08 | 0.98 | 1.91 | 1.03 |
| 0.47 | 2.43 | 1.10 | 2.15 | 1.05 | 2.08 | 0.97 | 1.93 | 1.03 |
| 0.48 | 2.45 | 1.06 | 2.18 | 1.01 | 2.08 | 0.97 | 1.96 | 1.02 |
| 0.49 | 2.47 | 1.02 | 2.19 | 0.97 | 2.08 | 0.97 | 1.98 | 1.00 |
| 0.5 | 2.49 | 0.99 | 2.20 | 0.92 | 2.09 | 0.97 | 2.01 | 0.98 |
| 0.51 | 2.50 | 0.95 | 2.20 | 0.88 | 2.10 | 0.98 | 2.03 | 0.96 |
| 0.52 | 2.51 | 0.92 | 2.20 | 0.84 | 2.12 | 0.99 | 2.05 | 0.94 |
| 0.53 | 2.51 | 0.89 | 2.18 | 0.80 | 2.14 | 1.00 | 2.06 | 0.91 |
| 0.54 | 2.51 | 0.86 | 2.17 | 0.76 | 2.16 | 1.00 | 2.08 | 0.88 |
| 0.55 | 2.51 | 0.83 | 2.14 | 0.73 | 2.19 | 1.00 | 2.08 | 0.85 |
| 0.56 | 2.51 | 0.81 | 2.11 | 0.70 | 2.21 | 1.00 | 2.08 | 0.81 |
| 0.57 | 2.51 | 0.79 | 2.08 | 0.68 | 2.24 | 1.00 | 2.08 | 0.78 |
| 0.58 | 2.50 | 0.77 | 2.05 | 0.67 | 2.27 | 0.99 | 2.06 | 0.75 |
| 0.59 | 2.50 | 0.75 | 2.01 | 0.66 | 2.29 | 0.97 | 2.05 | 0.72 |
| 0.6 | 2.49 | 0.74 | 1.98 | 0.65 | 2.31 | 0.96 | 2.03 | 0.69 |
| 0.61 | 2.49 | 0.73 | 1.94 | 0.65 | 2.33 | 0.94 | 2.00 | 0.66 |
| 0.62 | 2.48 | 0.72 | 1.90 | 0.66 | 2.34 | 0.93 | 1.98 | 0.64 |
| 0.63 | 2.48 | 0.71 | 1.87 | 0.67 | 2.35 | 0.91 | 1.95 | 0.63 |
| 0.64 | 2.47 | 0.70 | 1.83 | 0.68 | 2.36 | 0.89 | 1.92 | 0.61 |
| 0.65 | 2.47 | 0.70 | 1.80 | 0.71 | 2.37 | 0.88 | 1.88 | 0.60 |
| 0.66 | 2.46 | 0.69 | 1.77 | 0.73 | 2.37 | 0.86 | 1.85 | 0.60 |
| 0.67 | 2.46 | 0.69 | 1.75 | 0.76 | 2.38 | 0.85 | 1.81 | 0.60 |
| 0.68 | 2.45 | 0.69 | 1.74 | 0.79 | 2.38 | 0.83 | 1.78 | 0.60 |
| 0.69 | 2.45 | 0.69 | 1.73 | 0.83 | 2.38 | 0.82 | 1.74 | 0.60 |
| 0.7 | 2.45 | 0.69 | 1.72 | 0.86 | 2.38 | 0.82 | 1.71 | 0.61 |
| 0.71 | 2.45 | 0.69 | 1.73 | 0.90 | 2.38 | 0.81 | 1.67 | 0.63 |
| 0.72 | 2.45 | 0.70 | 1.74 | 0.92 | 2.37 | 0.80 | 1.64 | 0.64 |
| 0.73 | 2.45 | 0.70 | 1.76 | 0.94 | 2.37 | 0.80 | 1.61 | 0.66 |
| 0.74 | 2.45 | 0.70 | 1.78 | 0.95 | 2.37 | 0.80 | 1.58 | 0.68 |
| 0.75 | 2.46 | 0.70 | 1.79 | 0.96 | 2.37 | 0.79 | 1.56 | 0.70 |
| 0.76 | 2.46 | 0.71 | 1.80 | 0.95 | 2.37 | 0.79 | 1.53 | 0.73 |
| 0.77 | 2.47 | 0.71 | 1.81 | 0.94 | 2.37 | 0.80 | 1.51 | 0.75 |
| 0.78 | 2.47 | 0.72 | 1.81 | 0.93 | 2.37 | 0.80 | 1.49 | 0.78 |
| 0.79 | 2.48 | 0.72 | 1.80 | 0.92 | 2.37 | 0.80 | 1.47 | 0.80 |
| 0.8 | 2.49 | 0.72 | 1.79 | 0.91 | 2.37 | 0.80 | 1.46 | 0.83 |
| 0.81 | 2.49 | 0.72 | 1.77 | 0.91 | 2.37 | 0.81 | 1.44 | 0.86 |
| 0.82 | 2.50 | 0.73 | 1.75 | 0.90 | 2.37 | 0.81 | 1.43 | 0.88 |
| 0.83 | 2.51 | 0.73 | 1.73 | 0.91 | 2.38 | 0.82 | 1.42 | 0.91 |
| 0.84 | 2.52 | 0.73 | 1.71 | 0.91 | 2.38 | 0.82 | 1.41 | 0.93 |
| 0.85 | 2.53 | 0.73 | 1.68 | 0.92 | 2.38 | 0.83 | 1.40 | 0.96 |
| 0.86 | 2.54 | 0.73 | 1.66 | 0.93 | 2.39 | 0.83 | 1.40 | 0.98 |
| 0.87 | 2.55 | 0.73 | 1.64 | 0.94 | 2.40 | 0.84 | 1.39 | 1.00 |
| 0.88 | 2.56 | 0.72 | 1.61 | 0.95 | 2.40 | 0.84 | 1.38 | 1.03 |
| 0.89 | 2.57 | 0.72 | 1.59 | 0.97 | 2.41 | 0.84 | 1.37 | 1.05 |
| 0.9 | 2.58 | 0.72 | 1.57 | 0.99 | 2.42 | 0.85 | 1.37 | 1.07 |
| 0.91 | 2.59 | 0.72 | 1.55 | 1.01 | 2.43 | 0.85 | 1.36 | 1.09 |
| 0.92 | 2.60 | 0.72 | 1.53 | 1.03 | 2.43 | 0.86 | 1.35 | 1.11 |
| 0.93 | 2.61 | 0.71 | 1.51 | 1.05 | 2.44 | 0.86 | 1.35 | 1.13 |
| 0.94 | 2.61 | 0.71 | 1.49 | 1.07 | 2.45 | 0.86 | 1.34 | 1.15 |
| 0.95 | 2.62 | 0.71 | 1.48 | 1.10 | 2.46 | 0.87 | 1.33 | 1.17 |
| 0.96 | 2.63 | 0.70 | 1.46 | 1.12 | 2.47 | 0.87 | 1.32 | 1.19 |
| 0.97 | 2.64 | 0.70 | 1.45 | 1.15 | 2.48 | 0.87 | 1.31 | 1.21 |
| 0.98 | 2.65 | 0.69 | 1.44 | 1.18 | 2.49 | 0.87 | 1.30 | 1.23 |
| 0.99 | 2.65 | 0.69 | 1.42 | 1.20 | 2.50 | 0.88 | 1.29 | 1.25 |
| 1 | 2.66 | 0.69 | 1.41 | 1.23 | 2.51 | 0.88 | 1.29 | 1.28 |
| 1.01 | 2.67 | 0.68 | 1.40 | 1.26 | 2.52 | 0.88 | 1.28 | 1.30 |
| 1.02 | 2.68 | 0.68 | 1.39 | 1.29 | 2.53 | 0.88 | 1.27 | 1.32 |
| 1.03 | 2.69 | 0.68 | 1.38 | 1.32 | 2.54 | 0.88 | 1.26 | 1.35 |
| 1.04 | 2.69 | 0.67 | 1.38 | 1.35 | 2.55 | 0.88 | 1.25 | 1.37 |
| 1.05 | 2.70 | 0.67 | 1.37 | 1.37 | 2.56 | 0.88 | 1.24 | 1.40 |
| 1.06 | 2.71 | 0.66 | 1.36 | 1.40 | 2.57 | 0.89 | 1.24 | 1.42 |
| 1.07 | 2.71 | 0.66 | 1.36 | 1.43 | 2.58 | 0.89 | 1.23 | 1.45 |
| 1.08 | 2.72 | 0.65 | 1.35 | 1.46 | 2.59 | 0.89 | 1.22 | 1.47 |
| 1.09 | 2.73 | 0.65 | 1.35 | 1.49 | 2.60 | 0.89 | 1.22 | 1.50 |
| 1.1 | 2.74 | 0.65 | 1.35 | 1.52 | 2.61 | 0.89 | 1.21 | 1.53 |
| 1.11 | 2.74 | 0.64 | 1.34 | 1.55 | 2.62 | 0.89 | 1.21 | 1.55 |
| 1.12 | 2.75 | 0.64 | 1.34 | 1.58 | 2.63 | 0.89 | 1.20 | 1.58 |
| 1.13 | 2.76 | 0.63 | 1.34 | 1.60 | 2.64 | 0.89 | 1.20 | 1.61 |
| 1.14 | 2.77 | 0.63 | 1.34 | 1.63 | 2.66 | 0.89 | 1.20 | 1.63 |
| 1.15 | 2.77 | 0.62 | 1.34 | 1.66 | 2.67 | 0.89 | 1.19 | 1.66 |
| 1.16 | 2.78 | 0.61 | 1.34 | 1.69 | 2.68 | 0.89 | 1.19 | 1.69 |
| 1.17 | 2.79 | 0.61 | 1.34 | 1.72 | 2.69 | 0.88 | 1.19 | 1.72 |
| 1.18 | 2.79 | 0.60 | 1.34 | 1.74 | 2.70 | 0.88 | 1.19 | 1.74 |
| 1.19 | 2.80 | 0.60 | 1.34 | 1.77 | 2.72 | 0.88 | 1.19 | 1.77 |
| 1.2 | 2.80 | 0.59 | 1.35 | 1.80 | 2.73 | 0.88 | 1.19 | 1.80 |
| 1.21 | 2.81 | 0.58 | 1.35 | 1.83 | 2.74 | 0.87 | 1.19 | 1.82 |
| 1.22 | 2.82 | 0.58 | 1.35 | 1.85 | 2.75 | 0.87 | 1.19 | 1.85 |
| 1.23 | 2.82 | 0.57 | 1.36 | 1.88 | 2.77 | 0.87 | 1.19 | 1.88 |
| 1.24 | 2.83 | 0.56 | 1.36 | 1.91 | 2.78 | 0.86 | 1.19 | 1.91 |
| 1.25 | 2.83 | 0.55 | 1.36 | 1.93 | 2.79 | 0.86 | 1.19 | 1.93 |
| 1.26 | 2.84 | 0.55 | 1.37 | 1.96 | 2.80 | 0.85 | 1.19 | 1.96 |
| 1.27 | 2.84 | 0.54 | 1.37 | 1.98 | 2.81 | 0.85 | 1.20 | 1.99 |
| 1.28 | 2.84 | 0.53 | 1.38 | 2.01 | 2.83 | 0.84 | 1.20 | 2.01 |
| 1.29 | 2.85 | 0.52 | 1.38 | 2.03 | 2.84 | 0.83 | 1.20 | 2.04 |
| 1.3 | 2.85 | 0.52 | 1.39 | 2.06 | 2.85 | 0.83 | 1.20 | 2.06 |
| 1.31 | 2.85 | 0.51 | 1.39 | 2.08 | 2.86 | 0.82 | 1.21 | 2.09 |
| 1.32 | 2.86 | 0.50 | 1.40 | 2.11 | 2.87 | 0.81 | 1.21 | 2.12 |
| 1.33 | 2.86 | 0.50 | 1.40 | 2.13 | 2.88 | 0.81 | 1.22 | 2.14 |
| 1.34 | 2.86 | 0.49 | 1.41 | 2.16 | 2.89 | 0.80 | 1.22 | 2.17 |
| 1.43 | 2.88 | 0.43 | 1.48 | 2.37 | 2.96 | 0.72 | 1.27 | 2.39 |
| 1.44 | 2.88 | 0.42 | 1.48 | 2.39 | 2.97 | 0.71 | 1.27 | 2.41 |
| 1.45 | 2.88 | 0.42 | 1.49 | 2.41 | 2.97 | 0.70 | 1.28 | 2.44 |
| 1.46 | 2.88 | 0.41 | 1.50 | 2.44 | 2.98 | 0.69 | 1.29 | 2.46 |
| 1.47 | 2.88 | 0.40 | 1.51 | 2.46 | 2.98 | 0.68 | 1.29 | 2.48 |
| 1.48 | 2.88 | 0.40 | 1.52 | 2.48 | 2.98 | 0.67 | 1.30 | 2.51 |
| 1.49 | 2.88 | 0.39 | 1.52 | 2.50 | 2.99 | 0.66 | 1.31 | 2.53 |
| 1.5 | 2.88 | 0.39 | 1.53 | 2.52 | 2.99 | 0.65 | 1.31 | 2.55 |
| 1.51 | 2.88 | 0.38 | 1.54 | 2.55 | 3.00 | 0.64 | 1.32 | 2.57 |
| 1.52 | 2.88 | 0.38 | 1.55 | 2.57 | 3.00 | 0.63 | 1.33 | 2.60 |
| 1.53 | 2.88 | 0.37 | 1.56 | 2.59 | 3.00 | 0.62 | 1.34 | 2.62 |
| 1.54 | 2.88 | 0.37 | 1.57 | 2.61 | 3.00 | 0.61 | 1.34 | 2.64 |
| 1.55 | 2.88 | 0.36 | 1.58 | 2.63 | 3.01 | 0.61 | 1.35 | 2.66 |
| 1.56 | 2.88 | 0.36 | 1.59 | 2.65 | 3.01 | 0.60 | 1.36 | 2.69 |
| 1.57 | 2.88 | 0.35 | 1.60 | 2.67 | 3.01 | 0.59 | 1.37 | 2.71 |
| 1.58 | 2.88 | 0.35 | 1.61 | 2.69 | 3.01 | 0.58 | 1.37 | 2.73 |
| 1.59 | 2.88 | 0.34 | 1.62 | 2.71 | 3.01 | 0.57 | 1.38 | 2.75 |
| 1.6 | 2.88 | 0.34 | 1.63 | 2.73 | 3.02 | 0.56 | 1.39 | 2.77 |
| 1.61 | 2.88 | 0.34 | 1.63 | 2.75 | 3.02 | 0.56 | 1.40 | 2.79 |
| 1.62 | 2.87 | 0.33 | 1.64 | 2.77 | 3.02 | 0.55 | 1.41 | 2.81 |
| 1.63 | 2.87 | 0.33 | 1.65 | 2.79 | 3.02 | 0.54 | 1.42 | 2.84 |
| 1.64 | 2.87 | 0.32 | 1.66 | 2.81 | 3.02 | 0.53 | 1.42 | 2.86 |
| 1.65 | 2.87 | 0.32 | 1.67 | 2.83 | 3.02 | 0.53 | 1.43 | 2.88 |
| 1.66 | 2.87 | 0.32 | 1.68 | 2.85 | 3.02 | 0.52 | 1.44 | 2.90 |
| 1.67 | 2.87 | 0.31 | 1.69 | 2.87 | 3.02 | 0.51 | 1.45 | 2.92 |
| 1.68 | 2.87 | 0.31 | 1.70 | 2.89 | 3.02 | 0.51 | 1.46 | 2.94 |
| 1.69 | 2.87 | 0.31 | 1.71 | 2.91 | 3.02 | 0.50 | 1.47 | 2.96 |
| 1.7 | 2.87 | 0.30 | 1.72 | 2.93 | 3.02 | 0.49 | 1.48 | 2.98 |
| 1.71 | 2.87 | 0.30 | 1.73 | 2.94 | 3.02 | 0.49 | 1.48 | 3.00 |
| 1.72 | 2.87 | 0.30 | 1.74 | 2.96 | 3.02 | 0.48 | 1.49 | 3.02 |
| 1.73 | 2.87 | 0.29 | 1.75 | 2.98 | 3.02 | 0.48 | 1.50 | 3.04 |
| 1.74 | 2.87 | 0.29 | 1.77 | 3.00 | 3.02 | 0.47 | 1.51 | 3.06 |
| 1.75 | 2.87 | 0.29 | 1.78 | 3.02 | 3.02 | 0.46 | 1.52 | 3.08 |
| 1.76 | 2.86 | 0.28 | 1.79 | 3.04 | 3.02 | 0.46 | 1.53 | 3.10 |
| 1.77 | 2.86 | 0.28 | 1.80 | 3.05 | 3.02 | 0.45 | 1.54 | 3.12 |
| 1.78 | 2.86 | 0.28 | 1.81 | 3.07 | 3.02 | 0.45 | 1.55 | 3.14 |
| 1.79 | 2.86 | 0.28 | 1.82 | 3.09 | 3.02 | 0.44 | 1.56 | 3.16 |
| 1.8 | 2.86 | 0.27 | 1.83 | 3.11 | 3.02 | 0.44 | 1.57 | 3.17 |
| 1.81 | 2.86 | 0.27 | 1.84 | 3.12 | 3.02 | 0.43 | 1.58 | 3.19 |
| 1.82 | 2.86 | 0.27 | 1.85 | 3.14 | 3.02 | 0.43 | 1.58 | 3.21 |
| 1.83 | 2.86 | 0.27 | 1.86 | 3.16 | 3.02 | 0.42 | 1.59 | 3.23 |
| 1.84 | 2.86 | 0.26 | 1.87 | 3.18 | 3.02 | 0.42 | 1.60 | 3.25 |
| 1.85 | 2.86 | 0.26 | 1.88 | 3.19 | 3.02 | 0.41 | 1.61 | 3.27 |
| 1.86 | 2.86 | 0.26 | 1.89 | 3.21 | 3.02 | 0.41 | 1.62 | 3.29 |
| 1.87 | 2.86 | 0.26 | 1.90 | 3.23 | 3.01 | 0.40 | 1.63 | 3.30 |
| 1.88 | 2.86 | 0.25 | 1.91 | 3.24 | 3.01 | 0.40 | 1.64 | 3.32 |
| 1.89 | 2.86 | 0.25 | 1.93 | 3.26 | 3.01 | 0.40 | 1.65 | 3.34 |
| 1.9 | 2.85 | 0.25 | 1.94 | 3.28 | 3.01 | 0.39 | 1.66 | 3.36 |
| 1.91 | 2.85 | 0.25 | 1.95 | 3.29 | 3.01 | 0.39 | 1.67 | 3.38 |
| 1.92 | 2.85 | 0.24 | 1.96 | 3.31 | 3.01 | 0.38 | 1.68 | 3.39 |
| 1.93 | 2.85 | 0.24 | 1.97 | 3.33 | 3.01 | 0.38 | 1.69 | 3.41 |
| 1.94 | 2.85 | 0.24 | 1.98 | 3.34 | 3.01 | 0.38 | 1.70 | 3.43 |
| 1.95 | 2.85 | 0.24 | 1.99 | 3.36 | 3.01 | 0.37 | 1.71 | 3.45 |
| 1.96 | 2.85 | 0.24 | 2.00 | 3.37 | 3.01 | 0.37 | 1.72 | 3.46 |
| 1.97 | 2.85 | 0.23 | 2.01 | 3.39 | 3.01 | 0.36 | 1.73 | 3.48 |
| 1.98 | 2.85 | 0.23 | 2.03 | 3.41 | 3.01 | 0.36 | 1.74 | 3.50 |
| 1.99 | 2.85 | 0.23 | 2.04 | 3.42 | 3.00 | 0.36 | 1.75 | 3.52 |
| 2 | 2.85 | 0.23 | 2.05 | 3.44 | 3.00 | 0.35 | 1.76 | 3.53 |
| 2.1 | 2.82 | 0.19 | 2.20 | 3.58 | 2.99 | 0.31 | 1.89 | 3.76 |
| 2.2 | 2.81 | 0.17 | 2.31 | 3.71 | 2.98 | 0.29 | 1.99 | 3.92 |
| 2.3 | 2.80 | 0.16 | 2.42 | 3.84 | 2.97 | 0.27 | 2.09 | 4.07 |
| 2.4 | 2.79 | 0.15 | 2.52 | 3.96 | 2.96 | 0.25 | 2.19 | 4.22 |
| 2.5 | 2.78 | 0.14 | 2.63 | 4.08 | 2.95 | 0.24 | 2.29 | 4.37 |
| 2.6 | 2.77 | 0.13 | 2.74 | 4.19 | 2.94 | 0.22 | 2.40 | 4.51 |
| 2.7 | 2.77 | 0.12 | 2.85 | 4.30 | 2.93 | 0.21 | 2.50 | 4.64 |
| 2.8 | 2.76 | 0.12 | 2.95 | 4.41 | 2.92 | 0.20 | 2.61 | 4.77 |
| 2.9 | 2.75 | 0.11 | 3.06 | 4.51 | 2.91 | 0.19 | 2.72 | 4.90 |
| 3 | 2.75 | 0.11 | 3.17 | 4.62 | 2.91 | 0.18 | 2.82 | 5.02 |
| 3.1 | 2.74 | 0.10 | 3.28 | 4.71 | 2.90 | 0.17 | 2.93 | 5.14 |
| 3.2 | 2.74 | 0.10 | 3.38 | 4.81 | 2.89 | 0.17 | 3.04 | 5.25 |
| 3.3 | 2.73 | 0.10 | 3.49 | 4.90 | 2.88 | 0.16 | 3.15 | 5.37 |
| 3.4 | 2.73 | 0.09 | 3.60 | 4.99 | 2.87 | 0.15 | 3.25 | 5.48 |
| 3.5 | 2.72 | 0.09 | 3.71 | 5.08 | 2.87 | 0.15 | 3.36 | 5.58 |
| 3.6 | 2.72 | 0.09 | 3.82 | 5.16 | 2.86 | 0.14 | 3.47 | 5.69 |
| 3.7 | 2.72 | 0.09 | 3.94 | 5.24 | 2.85 | 0.14 | 3.58 | 5.79 |
| 3.8 | 2.71 | 0.08 | 4.05 | 5.32 | 2.85 | 0.14 | 3.69 | 5.89 |
| 3.9 | 2.71 | 0.08 | 4.16 | 5.39 | 2.84 | 0.13 | 3.80 | 5.99 |
| 4 | 2.70 | 0.08 | 4.27 | 5.46 | 2.84 | 0.13 | 3.91 | 6.09 |
| 4.1 | 2.70 | 0.08 | 4.39 | 5.52 | 2.83 | 0.13 | 4.02 | 6.18 |
| 4.2 | 2.69 | 0.08 | 4.50 | 5.59 | 2.82 | 0.12 | 4.13 | 6.27 |
| 4.3 | 2.69 | 0.08 | 4.61 | 5.65 | 2.82 | 0.12 | 4.25 | 6.36 |
| 4.4 | 2.68 | 0.08 | 4.72 | 5.70 | 2.81 | 0.12 | 4.36 | 6.45 |
| 4.5 | 2.67 | 0.07 | 4.84 | 5.75 | 2.81 | 0.12 | 4.48 | 6.54 |
| 4.6 | 2.67 | 0.07 | 4.95 | 5.80 | 2.80 | 0.11 | 4.60 | 6.62 |
| 4.7 | 2.66 | 0.07 | 5.06 | 5.84 | 2.80 | 0.11 | 4.72 | 6.70 |
| 4.8 | 2.66 | 0.07 | 5.16 | 5.88 | 2.79 | 0.11 | 4.84 | 6.77 |
| 4.9 | 2.65 | 0.07 | 5.27 | 5.92 | 2.79 | 0.11 | 4.96 | 6.85 |
| 5 | 2.65 | 0.07 | 5.38 | 5.96 | 2.78 | 0.11 | 5.08 | 6.91 |
| 5.1 | 2.64 | 0.07 | 5.48 | 5.99 | 2.77 | 0.10 | 5.20 | 6.98 |
| 5.2 | 2.63 | 0.07 | 5.59 | 6.02 | 2.77 | 0.10 | 5.33 | 7.04 |
| 5.3 | 2.63 | 0.07 | 5.69 | 6.04 | 2.76 | 0.10 | 5.45 | 7.10 |
| 5.4 | 2.62 | 0.07 | 5.79 | 6.07 | 2.76 | 0.10 | 5.58 | 7.15 |
| 5.5 | 2.61 | 0.07 | 5.89 | 6.09 | 2.75 | 0.10 | 5.70 | 7.20 |
| 5.6 | 2.61 | 0.07 | 5.98 | 6.11 | 2.75 | 0.10 | 5.83 | 7.24 |
| 5.7 | 2.60 | 0.07 | 6.08 | 6.13 | 2.74 | 0.10 | 5.95 | 7.28 |
| 5.8 | 2.59 | 0.07 | 6.17 | 6.14 | 2.74 | 0.10 | 6.07 | 7.31 |
| 5.9 | 2.58 | 0.07 | 6.26 | 6.15 | 2.73 | 0.10 | 6.19 | 7.35 |
| 6 | 2.58 | 0.07 | 6.35 | 6.17 | 2.72 | 0.09 | 6.31 | 7.37 |
| 6.1 | 2.57 | 0.07 | 6.43 | 6.18 | 2.72 | 0.09 | 6.42 | 7.40 |
| 6.2 | 2.56 | 0.07 | 6.52 | 6.19 | 2.71 | 0.09 | 6.54 | 7.42 |
| 6.3 | 2.55 | 0.07 | 6.60 | 6.19 | 2.71 | 0.09 | 6.65 | 7.44 |
| 6.4 | 2.55 | 0.07 | 6.68 | 6.20 | 2.70 | 0.09 | 6.76 | 7.45 |
| 6.5 | 2.54 | 0.07 | 6.76 | 6.21 | 2.69 | 0.09 | 6.86 | 7.46 |
| 6.6 | 2.53 | 0.07 | 6.84 | 6.21 | 2.69 | 0.09 | 6.97 | 7.48 |
| 6.7 | 2.52 | 0.07 | 6.91 | 6.22 | 2.68 | 0.09 | 7.07 | 7.48 |
| 6.8 | 2.51 | 0.07 | 6.99 | 6.22 | 2.67 | 0.09 | 7.17 | 7.49 |
| 6.9 | 2.50 | 0.07 | 7.06 | 6.22 | 2.67 | 0.09 | 7.26 | 7.50 |
| 7 | 2.49 | 0.07 | 7.14 | 6.23 | 2.66 | 0.09 | 7.36 | 7.51 |
| 7.1 | 2.48 | 0.07 | 7.21 | 6.23 | 2.65 | 0.09 | 7.45 | 7.51 |
| 7.2 | 2.47 | 0.07 | 7.28 | 6.23 | 2.65 | 0.09 | 7.54 | 7.51 |
| 7.3 | 2.46 | 0.07 | 7.36 | 6.23 | 2.64 | 0.09 | 7.63 | 7.52 |
| 7.4 | 2.45 | 0.08 | 7.43 | 6.23 | 2.63 | 0.09 | 7.72 | 7.52 |
| 7.5 | 2.44 | 0.08 | 7.50 | 6.23 | 2.63 | 0.09 | 7.81 | 7.52 |
| 7.6 | 2.43 | 0.08 | 7.57 | 6.23 | 2.62 | 0.09 | 7.89 | 7.53 |
| 7.7 | 2.42 | 0.08 | 7.65 | 6.23 | 2.61 | 0.09 | 7.98 | 7.53 |
| 7.8 | 2.40 | 0.08 | 7.72 | 6.22 | 2.60 | 0.09 | 8.07 | 7.53 |
| 7.9 | 2.39 | 0.08 | 7.80 | 6.21 | 2.59 | 0.09 | 8.16 | 7.53 |
| 8 | 2.38 | 0.08 | 7.88 | 6.20 | 2.58 | 0.09 | 8.25 | 7.53 |
| 8.1 | 2.37 | 0.08 | 7.97 | 6.19 | 2.58 | 0.09 | 8.34 | 7.53 |
| 8.2 | 2.35 | 0.08 | 8.06 | 6.16 | 2.57 | 0.09 | 8.44 | 7.52 |
| 8.3 | 2.34 | 0.09 | 8.15 | 6.13 | 2.56 | 0.10 | 8.54 | 7.51 |
| 8.4 | 2.33 | 0.09 | 8.23 | 6.08 | 2.55 | 0.10 | 8.65 | 7.50 |
| 8.5 | 2.31 | 0.09 | 8.32 | 6.01 | 2.54 | 0.10 | 8.76 | 7.47 |
| 8.6 | 2.30 | 0.09 | 8.39 | 5.94 | 2.53 | 0.10 | 8.88 | 7.43 |
| 8.7 | 2.28 | 0.09 | 8.45 | 5.84 | 2.52 | 0.10 | 9.00 | 7.37 |
| 8.8 | 2.27 | 0.10 | 8.49 | 5.74 | 2.51 | 0.10 | 9.12 | 7.29 |
| 8.9 | 2.25 | 0.10 | 8.51 | 5.63 | 2.50 | 0.10 | 9.22 | 7.17 |
| 9 | 2.24 | 0.10 | 8.50 | 5.52 | 2.49 | 0.10 | 9.30 | 7.03 |
| 9.1 | 2.22 | 0.10 | 8.46 | 5.43 | 2.48 | 0.11 | 9.34 | 6.86 |
| 9.2 | 2.20 | 0.11 | 8.41 | 5.34 | 2.46 | 0.11 | 9.33 | 6.70 |
| 9.3 | 2.18 | 0.11 | 8.33 | 5.28 | 2.45 | 0.11 | 9.27 | 6.55 |
| 9.4 | 2.17 | 0.11 | 8.25 | 5.24 | 2.44 | 0.11 | 9.17 | 6.43 |
| 9.5 | 2.15 | 0.12 | 8.16 | 5.22 | 2.43 | 0.12 | 9.04 | 6.35 |
| 9.6 | 2.13 | 0.12 | 8.08 | 5.22 | 2.41 | 0.12 | 8.91 | 6.32 |
| 9.7 | 2.11 | 0.12 | 8.00 | 5.24 | 2.40 | 0.12 | 8.78 | 6.33 |
| 9.8 | 2.09 | 0.13 | 7.92 | 5.28 | 2.39 | 0.12 | 8.67 | 6.38 |
| 9.9 | 2.07 | 0.13 | 7.86 | 5.33 | 2.37 | 0.13 | 8.58 | 6.45 |
| 10 | 2.04 | 0.14 | 7.81 | 5.39 | 2.36 | 0.13 | 8.51 | 6.53 |
| 10.1 | 2.02 | 0.14 | 7.78 | 5.45 | 2.34 | 0.13 | 8.47 | 6.62 |
| 10.2 | 2.00 | 0.15 | 7.75 | 5.52 | 2.32 | 0.14 | 8.45 | 6.71 |
| 10.3 | 1.97 | 0.15 | 7.73 | 5.59 | 2.31 | 0.14 | 8.43 | 6.79 |
| 10.4 | 1.95 | 0.16 | 7.72 | 5.66 | 2.29 | 0.15 | 8.43 | 6.88 |
| 10.5 | 1.92 | 0.17 | 7.71 | 5.73 | 2.27 | 0.15 | 8.44 | 6.96 |
| 10.6 | 1.89 | 0.17 | 7.71 | 5.79 | 2.25 | 0.16 | 8.46 | 7.03 |
| 10.7 | 1.86 | 0.18 | 7.72 | 5.86 | 2.23 | 0.17 | 8.48 | 7.11 |
| 10.8 | 1.83 | 0.19 | 7.73 | 5.92 | 2.21 | 0.17 | 8.50 | 7.18 |
| 10.9 | 1.80 | 0.20 | 7.75 | 5.98 | 2.19 | 0.18 | 8.53 | 7.24 |
| 11 | 1.77 | 0.21 | 7.76 | 6.04 | 2.17 | 0.19 | 8.56 | 7.31 |
| 11.1 | 1.74 | 0.23 | 7.78 | 6.10 | 2.15 | 0.20 | 8.59 | 7.37 |
| 11.2 | 1.70 | 0.24 | 7.80 | 6.16 | 2.13 | 0.21 | 8.63 | 7.43 |
| 11.3 | 1.67 | 0.25 | 7.83 | 6.22 | 2.10 | 0.22 | 8.67 | 7.49 |
| 11.4 | 1.63 | 0.27 | 7.85 | 6.27 | 2.08 | 0.23 | 8.71 | 7.54 |
| 11.5 | 1.59 | 0.29 | 7.88 | 6.33 | 2.05 | 0.24 | 8.75 | 7.60 |
| 11.6 | 1.55 | 0.31 | 7.91 | 6.38 | 2.03 | 0.26 | 8.79 | 7.65 |
| 11.7 | 1.50 | 0.33 | 7.94 | 6.43 | 2.00 | 0.27 | 8.83 | 7.70 |
| 11.8 | 1.46 | 0.36 | 7.97 | 6.48 | 1.97 | 0.29 | 8.88 | 7.76 |
| 11.9 | 1.41 | 0.39 | 8.01 | 6.53 | 1.94 | 0.31 | 8.93 | 7.80 |
| 12 | 1.36 | 0.42 | 8.04 | 6.58 | 1.91 | 0.33 | 8.97 | 7.85 |
| 12.1 | 1.31 | 0.46 | 8.08 | 6.63 | 1.88 | 0.35 | 9.02 | 7.90 |
| 12.2 | 1.26 | 0.51 | 8.12 | 6.67 | 1.84 | 0.38 | 9.07 | 7.94 |
| 12.3 | 1.21 | 0.56 | 8.16 | 6.72 | 1.81 | 0.41 | 9.12 | 7.98 |
| 12.4 | 1.16 | 0.61 | 8.20 | 6.76 | 1.78 | 0.44 | 9.17 | 8.03 |
| 12.5 | 1.12 | 0.68 | 8.24 | 6.81 | 1.74 | 0.48 | 9.22 | 8.07 |
| 12.6 | 1.07 | 0.75 | 8.28 | 6.85 | 1.71 | 0.52 | 9.27 | 8.11 |
| 12.7 | 1.03 | 0.83 | 8.32 | 6.89 | 1.68 | 0.56 | 9.33 | 8.14 |
| 12.8 | 1.00 | 0.91 | 8.36 | 6.93 | 1.64 | 0.61 | 9.38 | 8.18 |
| 12.9 | 0.98 | 0.99 | 8.40 | 6.97 | 1.62 | 0.67 | 9.43 | 8.22 |
| 13 | 0.96 | 1.08 | 8.45 | 7.00 | 1.59 | 0.72 | 9.48 | 8.25 |
| 13.1 | 0.95 | 1.17 | 8.49 | 7.04 | 1.56 | 0.79 | 9.54 | 8.28 |
| 13.2 | 0.94 | 1.27 | 8.54 | 7.07 | 1.55 | 0.85 | 9.59 | 8.32 |
| 13.3 | 0.94 | 1.36 | 8.58 | 7.11 | 1.53 | 0.92 | 9.65 | 8.35 |
| 13.4 | 0.95 | 1.45 | 8.63 | 7.14 | 1.52 | 1.00 | 9.70 | 8.38 |
| 13.5 | 0.97 | 1.54 | 8.67 | 7.17 | 1.52 | 1.07 | 9.76 | 8.41 |
| 13.6 | 0.99 | 1.63 | 8.72 | 7.20 | 1.53 | 1.15 | 9.81 | 8.43 |
| 13.7 | 1.01 | 1.72 | 8.76 | 7.23 | 1.54 | 1.22 | 9.86 | 8.46 |
| 13.8 | 1.04 | 1.81 | 8.81 | 7.26 | 1.55 | 1.30 | 9.92 | 8.49 |
| 13.9 | 1.08 | 1.90 | 8.86 | 7.29 | 1.57 | 1.37 | 9.97 | 8.51 |
| 14 | 1.12 | 1.98 | 8.90 | 7.32 | 1.60 | 1.44 | 10.03 | 8.53 |
| 14.1 | 1.17 | 2.06 | 8.95 | 7.34 | 1.63 | 1.51 | 10.08 | 8.56 |
| 14.2 | 1.22 | 2.15 | 9.00 | 7.37 | 1.66 | 1.57 | 10.14 | 8.58 |
| 14.3 | 1.27 | 2.22 | 9.04 | 7.39 | 1.70 | 1.63 | 10.19 | 8.60 |
| 14.4 | 1.33 | 2.30 | 9.09 | 7.41 | 1.74 | 1.69 | 10.24 | 8.62 |
| 14.5 | 1.40 | 2.37 | 9.14 | 7.43 | 1.78 | 1.75 | 10.30 | 8.64 |
| 14.6 | 1.47 | 2.44 | 9.18 | 7.45 | 1.81 | 1.81 | 10.35 | 8.66 |
| 14.7 | 1.55 | 2.51 | 9.23 | 7.47 | 1.85 | 1.87 | 10.40 | 8.67 |
| 14.8 | 1.63 | 2.57 | 9.28 | 7.49 | 1.89 | 1.93 | 10.46 | 8.69 |
| 14.9 | 1.71 | 2.64 | 9.32 | 7.51 | 1.94 | 2.00 | 10.51 | 8.70 |
| 15 | 1.81 | 2.70 | 9.37 | 7.53 | 1.99 | 2.07 | 10.56 | 8.72 |
| 15.1 | 1.92 | 2.76 | 9.41 | 7.54 | 2.06 | 2.15 | 10.61 | 8.73 |
| 15.2 | 2.04 | 2.81 | 9.46 | 7.56 | 2.14 | 2.23 | 10.66 | 8.75 |
| 15.3 | 2.17 | 2.84 | 9.50 | 7.57 | 2.24 | 2.30 | 10.72 | 8.76 |
| 15.4 | 2.31 | 2.85 | 9.55 | 7.59 | 2.36 | 2.35 | 10.77 | 8.77 |
| 15.5 | 2.45 | 2.84 | 9.59 | 7.60 | 2.49 | 2.37 | 10.82 | 8.78 |
| 15.6 | 2.59 | 2.81 | 9.64 | 7.61 | 2.61 | 2.37 | 10.87 | 8.79 |
| 15.7 | 2.70 | 2.76 | 9.68 | 7.62 | 2.70 | 2.36 | 10.92 | 8.80 |
| 15.8 | 2.79 | 2.71 | 9.72 | 7.63 | 2.78 | 2.35 | 10.97 | 8.81 |
| 15.9 | 2.88 | 2.68 | 9.77 | 7.64 | 2.85 | 2.37 | 11.01 | 8.82 |
| 16 | 2.97 | 2.66 | 9.81 | 7.65 | 2.95 | 2.40 | 11.06 | 8.83 |
| 16.1 | 3.10 | 2.64 | 9.85 | 7.66 | 3.08 | 2.44 | 11.11 | 8.83 |
| 16.2 | 3.26 | 2.59 | 9.89 | 7.67 | 3.26 | 2.46 | 11.16 | 8.84 |
| 16.3 | 3.44 | 2.48 | 9.93 | 7.68 | 3.48 | 2.41 | 11.20 | 8.85 |
| 16.4 | 3.59 | 2.31 | 9.97 | 7.68 | 3.68 | 2.29 | 11.25 | 8.85 |
| 16.5 | 3.68 | 2.09 | 10.01 | 7.69 | 3.84 | 2.10 | 11.30 | 8.86 |
| 16.6 | 3.70 | 1.86 | 10.05 | 7.70 | 3.93 | 1.88 | 11.34 | 8.86 |
| 16.7 | 3.67 | 1.65 | 10.09 | 7.70 | 3.96 | 1.67 | 11.39 | 8.87 |
| 16.8 | 3.59 | 1.47 | 10.13 | 7.71 | 3.93 | 1.48 | 11.43 | 8.87 |
| 16.9 | 3.49 | 1.33 | 10.17 | 7.71 | 3.87 | 1.31 | 11.47 | 8.87 |
| 17 | 3.39 | 1.22 | 10.20 | 7.72 | 3.80 | 1.19 | 11.52 | 8.88 |
| 17.1 | 3.28 | 1.14 | 10.24 | 7.72 | 3.71 | 1.08 | 11.56 | 8.88 |
| 17.2 | 3.18 | 1.07 | 10.28 | 7.73 | 3.63 | 1.00 | 11.60 | 8.88 |
| 17.3 | 3.07 | 1.03 | 10.31 | 7.73 | 3.54 | 0.94 | 11.64 | 8.88 |
| 17.4 | 2.97 | 0.99 | 10.35 | 7.73 | 3.45 | 0.89 | 11.69 | 8.89 |
| 17.5 | 2.87 | 0.97 | 10.38 | 7.73 | 3.37 | 0.86 | 11.73 | 8.89 |
| 17.6 | 2.77 | 0.96 | 10.42 | 7.74 | 3.29 | 0.84 | 11.77 | 8.89 |
| 17.7 | 2.67 | 0.96 | 10.45 | 7.74 | 3.21 | 0.82 | 11.81 | 8.89 |
| 17.8 | 2.57 | 0.97 | 10.48 | 7.74 | 3.13 | 0.82 | 11.85 | 8.89 |
| 17.9 | 2.47 | 0.99 | 10.51 | 7.74 | 3.06 | 0.82 | 11.88 | 8.89 |
| 18 | 2.38 | 1.02 | 10.55 | 7.74 | 2.99 | 0.83 | 11.92 | 8.89 |
| 18.1 | 2.29 | 1.07 | 10.58 | 7.74 | 2.91 | 0.85 | 11.96 | 8.89 |
| 18.2 | 2.20 | 1.13 | 10.61 | 7.74 | 2.84 | 0.88 | 12.00 | 8.89 |
| 18.3 | 2.11 | 1.21 | 10.64 | 7.75 | 2.78 | 0.92 | 12.03 | 8.89 |
| 18.4 | 2.05 | 1.31 | 10.67 | 7.75 | 2.71 | 0.97 | 12.07 | 8.89 |
| 18.5 | 1.99 | 1.43 | 10.70 | 7.75 | 2.65 | 1.03 | 12.11 | 8.89 |
| 18.6 | 1.97 | 1.57 | 10.73 | 7.75 | 2.60 | 1.10 | 12.14 | 8.89 |
| 18.7 | 1.99 | 1.72 | 10.75 | 7.75 | 2.56 | 1.19 | 12.18 | 8.89 |
| 18.8 | 2.05 | 1.86 | 10.78 | 7.75 | 2.54 | 1.29 | 12.21 | 8.88 |
| 18.9 | 2.15 | 1.98 | 10.81 | 7.75 | 2.53 | 1.40 | 12.24 | 8.88 |
| 19 | 2.28 | 2.06 | 10.84 | 7.75 | 2.55 | 1.51 | 12.28 | 8.88 |
| 19.1 | 2.41 | 2.09 | 10.86 | 7.75 | 2.59 | 1.63 | 12.31 | 8.88 |
| 19.2 | 2.54 | 2.09 | 10.89 | 7.75 | 2.67 | 1.74 | 12.34 | 8.88 |
| 19.3 | 2.65 | 2.08 | 10.91 | 7.75 | 2.78 | 1.84 | 12.37 | 8.88 |
| 19.4 | 2.76 | 2.04 | 10.94 | 7.75 | 2.91 | 1.90 | 12.41 | 8.87 |
| 19.5 | 2.87 | 1.98 | 10.96 | 7.75 | 3.06 | 1.93 | 12.44 | 8.87 |
| 19.6 | 2.97 | 1.89 | 10.99 | 7.75 | 3.21 | 1.92 | 12.47 | 8.87 |
| 19.7 | 3.04 | 1.76 | 11.01 | 7.75 | 3.36 | 1.87 | 12.50 | 8.87 |
| 19.8 | 3.06 | 1.63 | 11.04 | 7.75 | 3.48 | 1.78 | 12.53 | 8.87 |
| 19.9 | 3.05 | 1.50 | 11.06 | 7.75 | 3.56 | 1.68 | 12.56 | 8.86 |
| 20 | 3.00 | 1.38 | 11.08 | 7.75 | 3.62 | 1.56 | 12.59 | 8.86 |
| 20.1 | 2.93 | 1.29 | 11.10 | 7.75 | 3.64 | 1.44 | 12.62 | 8.86 |
| 20.2 | 2.85 | 1.21 | 11.12 | 7.75 | 3.64 | 1.33 | 12.64 | 8.86 |
| 20.3 | 2.76 | 1.16 | 11.15 | 7.75 | 3.62 | 1.23 | 12.67 | 8.85 |
| 20.4 | 2.67 | 1.13 | 11.17 | 7.75 | 3.58 | 1.15 | 12.70 | 8.85 |
| 20.5 | 2.58 | 1.12 | 11.19 | 7.75 | 3.54 | 1.08 | 12.73 | 8.85 |
| 20.6 | 2.50 | 1.12 | 11.21 | 7.75 | 3.49 | 1.03 | 12.75 | 8.85 |
| 20.7 | 2.42 | 1.14 | 11.23 | 7.75 | 3.44 | 0.98 | 12.78 | 8.85 |
| 20.8 | 2.36 | 1.17 | 11.25 | 7.75 | 3.38 | 0.94 | 12.80 | 8.84 |
| 20.9 | 2.30 | 1.21 | 11.27 | 7.75 | 3.33 | 0.92 | 12.83 | 8.84 |
| 21 | 2.27 | 1.25 | 11.29 | 7.75 | 3.28 | 0.90 | 12.86 | 8.84 |
| 21.1 | 2.25 | 1.29 | 11.30 | 7.75 | 3.22 | 0.88 | 12.88 | 8.84 |
| 21.2 | 2.25 | 1.31 | 11.32 | 7.75 | 3.17 | 0.87 | 12.90 | 8.83 |
| 21.3 | 2.23 | 1.31 | 11.34 | 7.75 | 3.12 | 0.87 | 12.93 | 8.83 |
| 21.4 | 2.21 | 1.30 | 11.36 | 7.75 | 3.08 | 0.87 | 12.95 | 8.83 |
| 21.5 | 2.17 | 1.29 | 11.38 | 7.75 | 3.03 | 0.88 | 12.98 | 8.83 |
| 21.6 | 2.12 | 1.28 | 11.39 | 7.75 | 2.98 | 0.89 | 13.00 | 8.83 |
| 21.7 | 2.06 | 1.29 | 11.41 | 7.75 | 2.94 | 0.90 | 13.02 | 8.82 |
| 21.8 | 2.00 | 1.31 | 11.43 | 7.75 | 2.90 | 0.91 | 13.05 | 8.82 |
| 21.9 | 1.93 | 1.34 | 11.44 | 7.76 | 2.86 | 0.93 | 13.07 | 8.82 |
| 22 | 1.87 | 1.39 | 11.46 | 7.76 | 2.82 | 0.95 | 13.09 | 8.82 |
| 22.1 | 1.82 | 1.44 | 11.47 | 7.76 | 2.78 | 0.98 | 13.11 | 8.82 |
| 22.2 | 1.77 | 1.50 | 11.49 | 7.76 | 2.74 | 1.00 | 13.13 | 8.82 |
| 22.3 | 1.73 | 1.56 | 11.51 | 7.76 | 2.71 | 1.03 | 13.15 | 8.81 |
| 22.4 | 1.70 | 1.63 | 11.52 | 7.76 | 2.68 | 1.07 | 13.17 | 8.81 |
| 22.5 | 1.68 | 1.70 | 11.54 | 7.77 | 2.64 | 1.10 | 13.20 | 8.81 |
| 22.6 | 1.67 | 1.78 | 11.55 | 7.77 | 2.62 | 1.14 | 13.22 | 8.81 |
| 22.7 | 1.66 | 1.85 | 11.57 | 7.77 | 2.59 | 1.18 | 13.24 | 8.81 |
| 22.8 | 1.66 | 1.92 | 11.58 | 7.77 | 2.56 | 1.23 | 13.26 | 8.81 |
| 22.9 | 1.67 | 1.99 | 11.59 | 7.78 | 2.54 | 1.27 | 13.28 | 8.81 |
| 23 | 1.69 | 2.06 | 11.61 | 7.78 | 2.52 | 1.32 | 13.30 | 8.81 |
| 23.1 | 1.72 | 2.13 | 11.62 | 7.78 | 2.51 | 1.38 | 13.31 | 8.81 |
| 23.2 | 1.75 | 2.19 | 11.64 | 7.78 | 2.50 | 1.43 | 13.33 | 8.80 |
| 23.3 | 1.79 | 2.24 | 11.65 | 7.79 | 2.49 | 1.49 | 13.35 | 8.80 |
| 23.4 | 1.83 | 2.29 | 11.66 | 7.79 | 2.49 | 1.55 | 13.37 | 8.80 |
| 23.5 | 1.88 | 2.33 | 11.68 | 7.79 | 2.50 | 1.62 | 13.39 | 8.80 |
| 23.6 | 1.93 | 2.36 | 11.69 | 7.80 | 2.51 | 1.68 | 13.41 | 8.80 |
| 23.7 | 1.98 | 2.39 | 11.70 | 7.80 | 2.54 | 1.74 | 13.42 | 8.80 |
| 23.8 | 2.03 | 2.40 | 11.71 | 7.80 | 2.57 | 1.80 | 13.44 | 8.80 |
| 23.9 | 2.09 | 2.41 | 11.73 | 7.81 | 2.61 | 1.86 | 13.46 | 8.80 |
| 24 | 2.14 | 2.40 | 11.74 | 7.81 | 2.65 | 1.91 | 13.48 | 8.80 |
| 24.1 | 2.18 | 2.39 | 11.75 | 7.82 | 2.70 | 1.95 | 13.49 | 8.80 |
| 24.2 | 2.22 | 2.37 | 11.76 | 7.82 | 2.76 | 1.99 | 13.51 | 8.80 |
| 24.3 | 2.25 | 2.34 | 11.77 | 7.82 | 2.83 | 2.02 | 13.53 | 8.80 |
| 24.4 | 2.27 | 2.31 | 11.79 | 7.83 | 2.89 | 2.03 | 13.54 | 8.80 |
| 24.5 | 2.29 | 2.28 | 11.80 | 7.83 | 2.96 | 2.03 | 13.56 | 8.80 |
| 24.6 | 2.29 | 2.25 | 11.81 | 7.84 | 3.02 | 2.02 | 13.58 | 8.80 |
| 24.7 | 2.29 | 2.21 | 11.82 | 7.84 | 3.08 | 2.00 | 13.59 | 8.80 |
| 24.8 | 2.27 | 2.18 | 11.83 | 7.85 | 3.12 | 1.98 | 13.61 | 8.81 |
| 24.9 | 2.25 | 2.15 | 11.84 | 7.85 | 3.16 | 1.94 | 13.62 | 8.81 |
| 25 | 2.23 | 2.13 | 11.85 | 7.86 | 3.19 | 1.90 | 13.64 | 8.81 |
| 25.1 | 2.19 | 2.11 | 11.87 | 7.86 | 3.21 | 1.85 | 13.65 | 8.81 |
| 25.2 | 2.16 | 2.10 | 11.88 | 7.87 | 3.22 | 1.81 | 13.67 | 8.81 |
| 25.3 | 2.12 | 2.09 | 11.89 | 7.87 | 3.22 | 1.77 | 13.68 | 8.81 |
| 25.4 | 2.07 | 2.09 | 11.90 | 7.88 | 3.20 | 1.73 | 13.70 | 8.81 |
| 25.5 | 2.02 | 2.10 | 11.91 | 7.88 | 3.18 | 1.69 | 13.71 | 8.81 |
| 25.6 | 1.98 | 2.11 | 11.92 | 7.89 | 3.15 | 1.66 | 13.73 | 8.82 |
| 25.7 | 1.93 | 2.13 | 11.93 | 7.90 | 3.12 | 1.63 | 13.74 | 8.82 |
| 25.8 | 1.88 | 2.15 | 11.94 | 7.90 | 3.08 | 1.62 | 13.76 | 8.82 |
| 25.9 | 1.84 | 2.19 | 11.95 | 7.91 | 3.03 | 1.60 | 13.77 | 8.82 |
| 26 | 1.79 | 2.22 | 11.96 | 7.91 | 2.99 | 1.60 | 13.79 | 8.82 |
| 26.1 | 1.75 | 2.26 | 11.97 | 7.92 | 2.94 | 1.60 | 13.80 | 8.82 |
| 26.2 | 1.72 | 2.31 | 11.98 | 7.93 | 2.88 | 1.61 | 13.81 | 8.83 |
| 26.3 | 1.68 | 2.36 | 11.99 | 7.93 | 2.83 | 1.63 | 13.83 | 8.83 |
| 26.4 | 1.65 | 2.41 | 12.00 | 7.94 | 2.77 | 1.65 | 13.84 | 8.83 |
| 26.5 | 1.62 | 2.47 | 12.01 | 7.95 | 2.72 | 1.69 | 13.85 | 8.83 |
| 26.6 | 1.60 | 2.53 | 12.02 | 7.95 | 2.67 | 1.73 | 13.87 | 8.84 |
| 26.7 | 1.57 | 2.59 | 12.03 | 7.96 | 2.61 | 1.77 | 13.88 | 8.84 |
| 26.8 | 1.56 | 2.66 | 12.04 | 7.97 | 2.56 | 1.83 | 13.89 | 8.84 |
| 26.9 | 1.54 | 2.72 | 12.05 | 7.97 | 2.52 | 1.89 | 13.91 | 8.84 |
| 27 | 1.53 | 2.79 | 12.06 | 7.98 | 2.47 | 1.96 | 13.92 | 8.85 |
| 27.1 | 1.53 | 2.86 | 12.07 | 7.99 | 2.43 | 2.04 | 13.93 | 8.85 |
| 27.2 | 1.52 | 2.93 | 12.08 | 8.00 | 2.40 | 2.12 | 13.95 | 8.85 |
| 27.3 | 1.52 | 3.00 | 12.09 | 8.00 | 2.37 | 2.21 | 13.96 | 8.86 |
| 27.4 | 1.53 | 3.08 | 12.10 | 8.01 | 2.35 | 2.30 | 13.97 | 8.86 |
| 27.5 | 1.54 | 3.15 | 12.11 | 8.02 | 2.33 | 2.40 | 13.98 | 8.86 |
| 27.6 | 1.55 | 3.22 | 12.11 | 8.03 | 2.32 | 2.51 | 13.99 | 8.87 |
| 27.7 | 1.56 | 3.30 | 12.12 | 8.03 | 2.32 | 2.61 | 14.01 | 8.87 |
| 27.8 | 1.58 | 3.37 | 12.13 | 8.04 | 2.33 | 2.73 | 14.02 | 8.87 |
| 27.9 | 1.60 | 3.45 | 12.14 | 8.05 | 2.34 | 2.84 | 14.03 | 8.88 |
| 28 | 1.62 | 3.52 | 12.15 | 8.06 | 2.37 | 2.96 | 14.04 | 8.88 |
| 28.1 | 1.65 | 3.60 | 12.16 | 8.07 | 2.40 | 3.08 | 14.05 | 8.88 |
| 28.2 | 1.68 | 3.67 | 12.17 | 8.07 | 2.44 | 3.20 | 14.07 | 8.89 |
| 28.3 | 1.72 | 3.75 | 12.18 | 8.08 | 2.49 | 3.32 | 14.08 | 8.89 |
| 28.4 | 1.76 | 3.82 | 12.19 | 8.09 | 2.55 | 3.44 | 14.09 | 8.89 |
| 28.5 | 1.80 | 3.90 | 12.20 | 8.10 | 2.62 | 3.57 | 14.10 | 8.90 |
| 28.6 | 1.85 | 3.98 | 12.21 | 8.11 | 2.70 | 3.69 | 14.11 | 8.90 |
| 28.7 | 1.90 | 4.05 | 12.21 | 8.12 | 2.79 | 3.81 | 14.12 | 8.91 |
| 28.8 | 1.95 | 4.13 | 12.22 | 8.12 | 2.90 | 3.93 | 14.14 | 8.91 |
| 28.9 | 2.01 | 4.20 | 12.23 | 8.13 | 3.01 | 4.05 | 14.15 | 8.92 |
| 29 | 2.08 | 4.27 | 12.24 | 8.14 | 3.14 | 4.16 | 14.16 | 8.92 |
| 29.1 | 2.15 | 4.35 | 12.25 | 8.15 | 3.28 | 4.26 | 14.17 | 8.93 |
| 29.2 | 2.22 | 4.42 | 12.26 | 8.16 | 3.44 | 4.36 | 14.18 | 8.93 |
| 29.3 | 2.30 | 4.48 | 12.27 | 8.17 | 3.61 | 4.45 | 14.19 | 8.93 |
| 29.4 | 2.39 | 4.55 | 12.28 | 8.18 | 3.79 | 4.53 | 14.20 | 8.94 |
| 29.5 | 2.48 | 4.62 | 12.28 | 8.19 | 3.98 | 4.59 | 14.21 | 8.94 |
| 29.6 | 2.58 | 4.68 | 12.29 | 8.20 | 4.18 | 4.64 | 14.22 | 8.95 |
| 29.7 | 2.68 | 4.73 | 12.30 | 8.20 | 4.40 | 4.67 | 14.23 | 8.95 |
| 29.8 | 2.79 | 4.78 | 12.31 | 8.21 | 4.62 | 4.67 | 14.25 | 8.96 |
| 29.9 | 2.91 | 4.83 | 12.32 | 8.22 | 4.85 | 4.66 | 14.26 | 8.96 |
| 30 | 3.03 | 4.87 | 12.33 | 8.23 | 5.07 | 4.62 | 14.27 | 8.97 |

| | FILM 3 | | | | FILM 4 | | | |
|---|---|---|---|---|---|---|---|---|
| WAVELENGTH (MICRON) | $n$ (insulating) | $\kappa$ (insulating) | $n$ (metallic) | $\kappa$ (metallic) | $n$ (insulating) | $\kappa$ (insulating) | $n$ (metallic) | $\kappa$ (metallic) |
| 0.3 | 1.48717 | 1.13661 | 1.48 | 1.02 | 1.89 | 1.26 | 1.88 | 1.12 |
| 0.31 | 1.53154 | 1.15216 | 1.48 | 1.03 | 1.92 | 1.33 | 1.87 | 1.13 |
| 0.32 | 1.57527 | 1.1662 | 1.50 | 1.06 | 1.97 | 1.40 | 1.87 | 1.16 |
| 0.33 | 1.61853 | 1.17885 | 1.52 | 1.09 | 2.04 | 1.45 | 1.88 | 1.19 |
| 0.34 | 1.66146 | 1.19022 | 1.55 | 1.11 | 2.13 | 1.48 | 1.90 | 1.24 |
| 0.35 | 1.70423 | 1.20039 | 1.59 | 1.14 | 2.22 | 1.50 | 1.95 | 1.28 |
| 0.36 | 1.74702 | 1.20942 | 1.62 | 1.15 | 2.31 | 1.50 | 2.01 | 1.31 |
| 0.37 | 1.79002 | 1.21733 | 1.66 | 1.17 | 2.40 | 1.49 | 2.07 | 1.33 |
| 0.38 | 1.8335 | 1.22412 | 1.70 | 1.18 | 2.48 | 1.46 | 2.14 | 1.33 |
| 0.39 | 1.87774 | 1.22975 | 1.74 | 1.19 | 2.56 | 1.42 | 2.22 | 1.32 |
| 0.4 | 1.92314 | 1.23412 | 1.78 | 1.20 | 2.63 | 1.38 | 2.28 | 1.30 |
| 0.41 | 1.9702 | 1.23702 | 1.83 | 1.20 | 2.68 | 1.33 | 2.34 | 1.27 |
| 0.42 | 2.01959 | 1.23799 | 1.87 | 1.20 | 2.73 | 1.27 | 2.40 | 1.22 |
| 0.43 | 2.07211 | 1.23614 | 1.91 | 1.19 | 2.77 | 1.22 | 2.44 | 1.17 |
| 0.44 | 2.12842 | 1.22988 | 1.96 | 1.18 | 2.80 | 1.17 | 2.47 | 1.12 |
| 0.45 | 2.18858 | 1.21687 | 2.00 | 1.16 | 2.81 | 1.12 | 2.50 | 1.07 |
| 0.46 | 2.25149 | 1.19429 | 2.05 | 1.13 | 2.83 | 1.08 | 2.51 | 1.01 |
| 0.47 | 2.31449 | 1.15948 | 2.09 | 1.10 | 2.84 | 1.05 | 2.52 | 0.96 |
| 0.48 | 2.37348 | 1.11087 | 2.13 | 1.05 | 2.86 | 1.02 | 2.52 | 0.91 |
| 0.49 | 2.42358 | 1.04871 | 2.16 | 1.00 | 2.88 | 0.99 | 2.51 | 0.86 |
| 0.5 | 2.46006 | 0.97527 | 2.17 | 0.93 | 2.91 | 0.96 | 2.50 | 0.82 |
| 0.51 | 2.47925 | 0.8946 | 2.16 | 0.86 | 2.94 | 0.92 | 2.48 | 0.78 |
| 0.52 | 2.47917 | 0.81183 | 2.12 | 0.80 | 2.97 | 0.87 | 2.46 | 0.74 |
| 0.53 | 2.4597 | 0.73241 | 2.07 | 0.76 | 2.98 | 0.81 | 2.44 | 0.71 |
| 0.54 | 2.42246 | 0.66137 | 2.01 | 0.74 | 2.99 | 0.75 | 2.41 | 0.68 |
| 0.55 | 2.37038 | 0.60286 | 1.96 | 0.74 | 2.98 | 0.70 | 2.38 | 0.66 |
| 0.56 | 2.30724 | 0.55987 | 1.92 | 0.75 | 2.96 | 0.65 | 2.35 | 0.64 |
| 0.57 | 2.23729 | 0.53418 | 1.88 | 0.77 | 2.93 | 0.60 | 2.32 | 0.62 |
| 0.58 | 2.16486 | 0.52646 | 1.86 | 0.80 | 2.91 | 0.57 | 2.29 | 0.61 |
| 0.59 | 2.0942 | 0.53633 | 1.85 | 0.83 | 2.88 | 0.55 | 2.25 | 0.60 |
| 0.6 | 2.02931 | 0.56248 | 1.85 | 0.85 | 2.85 | 0.53 | 2.22 | 0.59 |
| 0.61 | 1.97377 | 0.60258 | 1.85 | 0.88 | 2.83 | 0.52 | 2.19 | 0.59 |
| 0.62 | 1.93044 | 0.65341 | 1.86 | 0.89 | 2.80 | 0.51 | 2.15 | 0.58 |
| 0.63 | 1.9011 | 0.71095 | 1.88 | 0.90 | 2.78 | 0.50 | 2.12 | 0.58 |
| 0.64 | 1.88609 | 0.77082 | 1.89 | 0.89 | 2.76 | 0.50 | 2.08 | 0.59 |
| 0.65 | 1.88419 | 0.82893 | 1.90 | 0.88 | 2.75 | 0.50 | 2.05 | 0.59 |
| 0.66 | 1.89289 | 0.88216 | 1.90 | 0.87 | 2.74 | 0.50 | 2.02 | 0.60 |
| 0.67 | 1.90898 | 0.92878 | 1.90 | 0.85 | 2.73 | 0.51 | 1.98 | 0.61 |
| 0.68 | 1.92931 | 0.96854 | 1.89 | 0.84 | 2.72 | 0.51 | 1.95 | 0.62 |
| 0.69 | 1.95136 | 1.00236 | 1.87 | 0.83 | 2.71 | 0.51 | 1.92 | 0.63 |
| 0.7 | 1.97363 | 1.03172 | 1.85 | 0.82 | 2.70 | 0.52 | 1.88 | 0.65 |
| 0.71 | 1.99555 | 1.05815 | 1.83 | 0.81 | 2.70 | 0.52 | 1.85 | 0.67 |
| 0.72 | 2.01727 | 1.08286 | 1.81 | 0.81 | 2.70 | 0.53 | 1.82 | 0.69 |
| 0.73 | 2.03931 | 1.10651 | 1.79 | 0.81 | 2.70 | 0.53 | 1.79 | 0.72 |
| 0.74 | 2.06222 | 1.12928 | 1.77 | 0.82 | 2.69 | 0.54 | 1.76 | 0.75 |
| 0.75 | 2.08638 | 1.15103 | 1.75 | 0.82 | 2.69 | 0.54 | 1.74 | 0.78 |
| 0.76 | 2.11197 | 1.17138 | 1.73 | 0.83 | 2.70 | 0.54 | 1.71 | 0.81 |
| 0.77 | 2.13887 | 1.18996 | 1.71 | 0.84 | 2.70 | 0.55 | 1.69 | 0.85 |
| 0.78 | 2.16679 | 1.20646 | 1.68 | 0.85 | 2.70 | 0.55 | 1.68 | 0.88 |
| 0.79 | 2.19532 | 1.22074 | 1.66 | 0.86 | 2.70 | 0.56 | 1.67 | 0.92 |
| 0.8 | 2.22402 | 1.23282 | 1.64 | 0.87 | 2.71 | 0.56 | 1.66 | 0.95 |
| 0.81 | 2.25246 | 1.24285 | 1.62 | 0.88 | 2.71 | 0.57 | 1.66 | 0.98 |
| 0.82 | 2.28033 | 1.25113 | 1.60 | 0.90 | 2.72 | 0.57 | 1.65 | 1.01 |
| 0.83 | 2.30744 | 1.258 | 1.59 | 0.91 | 2.72 | 0.57 | 1.65 | 1.03 |
| 0.84 | 2.33369 | 1.26383 | 1.57 | 0.93 | 2.73 | 0.57 | 1.65 | 1.06 |
| 0.85 | 2.35915 | 1.26896 | 1.55 | 0.94 | 2.74 | 0.58 | 1.65 | 1.08 |
| 0.86 | 2.38392 | 1.27369 | 1.53 | 0.96 | 2.75 | 0.58 | 1.64 | 1.10 |
| 0.87 | 2.40822 | 1.27822 | 1.51 | 0.98 | 2.75 | 0.58 | 1.64 | 1.12 |
| 0.88 | 2.43225 | 1.28268 | 1.50 | 1.00 | 2.76 | 0.58 | 1.63 | 1.13 |
| 0.89 | 2.45627 | 1.28713 | 1.48 | 1.02 | 2.77 | 0.58 | 1.62 | 1.15 |
| 0.9 | 2.48049 | 1.29154 | 1.46 | 1.04 | 2.78 | 0.58 | 1.61 | 1.17 |
| 0.91 | 2.50511 | 1.29582 | 1.45 | 1.06 | 2.79 | 0.58 | 1.59 | 1.19 |
| 0.92 | 2.53027 | 1.29985 | 1.43 | 1.08 | 2.80 | 0.58 | 1.58 | 1.21 |
| 0.93 | 2.55607 | 1.30345 | 1.42 | 1.10 | 2.81 | 0.58 | 1.57 | 1.24 |
| 0.94 | 2.58257 | 1.30644 | 1.41 | 1.13 | 2.82 | 0.57 | 1.56 | 1.26 |
| 0.95 | 2.60977 | 1.30866 | 1.39 | 1.15 | 2.83 | 0.57 | 1.54 | 1.29 |
| 0.96 | 2.63762 | 1.30991 | 1.38 | 1.17 | 2.84 | 0.57 | 1.53 | 1.32 |
| 0.97 | 2.66606 | 1.31003 | 1.37 | 1.20 | 2.84 | 0.56 | 1.52 | 1.35 |
| 0.98 | 2.69496 | 1.30888 | 1.36 | 1.22 | 2.85 | 0.56 | 1.51 | 1.38 |
| 0.99 | 2.72418 | 1.30636 | 1.34 | 1.25 | 2.86 | 0.56 | 1.50 | 1.41 |
| 1 | 2.75357 | 1.30238 | 1.33 | 1.27 | 2.87 | 0.55 | 1.50 | 1.44 |
| 1.01 | 2.78297 | 1.29688 | 1.32 | 1.30 | 2.88 | 0.55 | 1.49 | 1.47 |
| 1.02 | 2.8122 | 1.28985 | 1.31 | 1.32 | 2.89 | 0.54 | 1.48 | 1.50 |
| 1.03 | 2.84108 | 1.28129 | 1.30 | 1.35 | 2.90 | 0.53 | 1.48 | 1.54 |
| 1.04 | 2.86946 | 1.27124 | 1.30 | 1.38 | 2.90 | 0.53 | 1.48 | 1.57 |
| 1.05 | 2.89719 | 1.25975 | 1.29 | 1.40 | 2.91 | 0.52 | 1.48 | 1.60 |
| 1.06 | 2.92412 | 1.24692 | 1.28 | 1.43 | 2.92 | 0.51 | 1.48 | 1.63 |
| 1.07 | 2.95014 | 1.23283 | 1.27 | 1.46 | 2.92 | 0.51 | 1.48 | 1.66 |
| 1.08 | 2.97514 | 1.21759 | 1.27 | 1.48 | 2.93 | 0.50 | 1.48 | 1.69 |
| 1.09 | 2.99904 | 1.20132 | 1.26 | 1.51 | 2.94 | 0.49 | 1.48 | 1.73 |
| 1.1 | 3.02177 | 1.18414 | 1.26 | 1.54 | 2.94 | 0.48 | 1.48 | 1.76 |
| 1.11 | 3.04328 | 1.16618 | 1.25 | 1.56 | 2.95 | 0.48 | 1.48 | 1.79 |
| 1.12 | 3.06355 | 1.14756 | 1.25 | 1.59 | 2.95 | 0.47 | 1.48 | 1.82 |
| 1.13 | 3.08256 | 1.12841 | 1.24 | 1.62 | 2.96 | 0.46 | 1.49 | 1.85 |
| 1.14 | 3.1003 | 1.10883 | 1.24 | 1.64 | 2.96 | 0.45 | 1.49 | 1.88 |
| 1.15 | 3.1168 | 1.08895 | 1.24 | 1.67 | 2.97 | 0.45 | 1.50 | 1.90 |
| 1.16 | 3.13208 | 1.06886 | 1.23 | 1.70 | 2.97 | 0.44 | 1.50 | 1.93 |
| 1.17 | 3.14616 | 1.04865 | 1.23 | 1.73 | 2.97 | 0.43 | 1.51 | 1.96 |
| 1.18 | 3.1591 | 1.02842 | 1.23 | 1.75 | 2.98 | 0.42 | 1.51 | 1.99 |
| 1.19 | 3.17092 | 1.00824 | 1.23 | 1.78 | 2.98 | 0.41 | 1.52 | 2.02 |
| 1.2 | 3.18168 | 0.98817 | 1.23 | 1.81 | 2.98 | 0.41 | 1.52 | 2.04 |
| 1.21 | 3.19144 | 0.96829 | 1.23 | 1.83 | 2.99 | 0.40 | 1.53 | 2.07 |
| 1.22 | 3.20024 | 0.94863 | 1.23 | 1.86 | 2.99 | 0.39 | 1.54 | 2.10 |
| 1.23 | 3.20815 | 0.92925 | 1.22 | 1.89 | 2.99 | 0.38 | 1.54 | 2.13 |
| 1.24 | 3.2152 | 0.91019 | 1.23 | 1.91 | 2.99 | 0.38 | 1.55 | 2.15 |
| 1.25 | 3.22147 | 0.89147 | 1.23 | 1.94 | 2.99 | 0.37 | 1.56 | 2.18 |
| 1.26 | 3.22701 | 0.87311 | 1.23 | 1.96 | 3.00 | 0.36 | 1.57 | 2.20 |
| 1.27 | 3.23185 | 0.85515 | 1.23 | 1.99 | 3.00 | 0.35 | 1.57 | 2.23 |
| 1.28 | 3.23606 | 0.83758 | 1.23 | 2.02 | 3.00 | 0.35 | 1.58 | 2.25 |
| 1.29 | 3.23968 | 0.82044 | 1.23 | 2.04 | 3.00 | 0.34 | 1.59 | 2.28 |
| 1.3 | 3.24276 | 0.80371 | 1.23 | 2.07 | 3.00 | 0.33 | 1.60 | 2.30 |
| 1.31 | 3.24533 | 0.78742 | 1.23 | 2.09 | 3.00 | 0.33 | 1.61 | 2.33 |
| 1.32 | 3.24744 | 0.77155 | 1.24 | 2.12 | 3.00 | 0.32 | 1.62 | 2.35 |
| 1.33 | 3.24912 | 0.75611 | 1.24 | 2.15 | 3.00 | 0.31 | 1.62 | 2.38 |
| 1.34 | 3.25042 | 0.74109 | 1.24 | 2.17 | 3.00 | 0.31 | 1.63 | 2.40 |
| 1.43 | 3.24936 | 0.62394 | 1.27 | 2.39 | 3.00 | 0.25 | 1.72 | 2.61 |
| 1.44 | 3.24824 | 0.61276 | 1.28 | 2.42 | 3.00 | 0.25 | 1.73 | 2.63 |
| 1.45 | 3.247 | 0.60191 | 1.28 | 2.44 | 3.00 | 0.24 | 1.74 | 2.65 |
| 1.46 | 3.24564 | 0.59138 | 1.29 | 2.47 | 3.00 | 0.24 | 1.75 | 2.67 |
| 1.47 | 3.24417 | 0.58116 | 1.29 | 2.49 | 2.99 | 0.23 | 1.76 | 2.69 |
| 1.48 | 3.2426 | 0.57124 | 1.30 | 2.51 | 2.99 | 0.23 | 1.77 | 2.71 |
| 1.49 | 3.24096 | 0.56161 | 1.30 | 2.54 | 2.99 | 0.22 | 1.78 | 2.73 |
| 1.5 | 3.23924 | 0.55227 | 1.31 | 2.56 | 2.99 | 0.22 | 1.79 | 2.75 |
| 1.51 | 3.23745 | 0.54319 | 1.31 | 2.58 | 2.99 | 0.21 | 1.80 | 2.77 |
| 1.52 | 3.23561 | 0.53437 | 1.32 | 2.61 | 2.99 | 0.21 | 1.81 | 2.79 |
| 1.53 | 3.23372 | 0.52581 | 1.33 | 2.63 | 2.99 | 0.20 | 1.82 | 2.81 |
| 1.54 | 3.23179 | 0.51749 | 1.33 | 2.65 | 2.99 | 0.20 | 1.83 | 2.83 |
| 1.55 | 3.22982 | 0.50941 | 1.34 | 2.67 | 2.98 | 0.20 | 1.84 | 2.85 |
| 1.56 | 3.22783 | 0.50156 | 1.34 | 2.70 | 2.98 | 0.19 | 1.85 | 2.87 |
| 1.57 | 3.22581 | 0.49392 | 1.35 | 2.72 | 2.98 | 0.19 | 1.86 | 2.89 |
| 1.58 | 3.22377 | 0.4865 | 1.36 | 2.74 | 2.98 | 0.18 | 1.87 | 2.91 |
| 1.59 | 3.22171 | 0.47928 | 1.36 | 2.77 | 2.98 | 0.18 | 1.89 | 2.93 |
| 1.6 | 3.21964 | 0.47226 | 1.37 | 2.79 | 2.98 | 0.18 | 1.90 | 2.95 |
| 1.61 | 3.21757 | 0.46543 | 1.38 | 2.81 | 2.98 | 0.17 | 1.91 | 2.97 |
| 1.62 | 3.21549 | 0.45878 | 1.38 | 2.83 | 2.97 | 0.17 | 1.92 | 2.99 |
| 1.63 | 3.21341 | 0.45231 | 1.39 | 2.85 | 2.97 | 0.17 | 1.93 | 3.01 |
| 1.64 | 3.21133 | 0.44601 | 1.40 | 2.88 | 2.97 | 0.16 | 1.94 | 3.03 |
| 1.65 | 3.20925 | 0.43987 | 1.40 | 2.90 | 2.97 | 0.16 | 1.95 | 3.04 |
| 1.66 | 3.20719 | 0.4339 | 1.41 | 2.92 | 2.97 | 0.16 | 1.96 | 3.06 |
| 1.67 | 3.20512 | 0.42808 | 1.42 | 2.94 | 2.97 | 0.15 | 1.97 | 3.08 |
| 1.68 | 3.20307 | 0.4224 | 1.42 | 2.96 | 2.96 | 0.15 | 1.98 | 3.10 |
| 1.69 | 3.20103 | 0.41687 | 1.43 | 2.98 | 2.96 | 0.15 | 1.99 | 3.12 |
| 1.7 | 3.19901 | 0.41148 | 1.44 | 3.00 | 2.96 | 0.14 | 2.01 | 3.13 |
| 1.71 | 3.19699 | 0.40623 | 1.45 | 3.03 | 2.96 | 0.14 | 2.02 | 3.15 |
| 1.72 | 3.19499 | 0.4011 | 1.45 | 3.05 | 2.96 | 0.14 | 2.03 | 3.17 |
| 1.73 | 3.19301 | 0.3961 | 1.46 | 3.07 | 2.96 | 0.13 | 2.04 | 3.19 |
| 1.74 | 3.19105 | 0.39122 | 1.47 | 3.09 | 2.95 | 0.13 | 2.05 | 3.20 |
| 1.75 | 3.18911 | 0.38645 | 1.48 | 3.11 | 2.95 | 0.13 | 2.06 | 3.22 |
| 1.76 | 3.18718 | 0.3818 | 1.48 | 3.13 | 2.95 | 0.13 | 2.07 | 3.24 |
| 1.77 | 3.18527 | 0.37726 | 1.49 | 3.15 | 2.95 | 0.12 | 2.08 | 3.26 |
| 1.78 | 3.18339 | 0.37282 | 1.50 | 3.17 | 2.95 | 0.12 | 2.09 | 3.27 |
| 1.79 | 3.18152 | 0.36849 | 1.51 | 3.19 | 2.95 | 0.12 | 2.10 | 3.29 |
| 1.8 | 3.17968 | 0.36426 | 1.51 | 3.21 | 2.95 | 0.12 | 2.12 | 3.31 |
| 1.81 | 3.17785 | 0.36012 | 1.52 | 3.23 | 2.94 | 0.12 | 2.13 | 3.32 |
| 1.82 | 3.17605 | 0.35607 | 1.53 | 3.25 | 2.94 | 0.11 | 2.14 | 3.34 |
| 1.83 | 3.17428 | 0.35212 | 1.54 | 3.27 | 2.94 | 0.11 | 2.15 | 3.35 |
| 1.84 | 3.17252 | 0.34825 | 1.55 | 3.29 | 2.94 | 0.11 | 2.16 | 3.37 |
| 1.85 | 3.17079 | 0.34447 | 1.55 | 3.31 | 2.94 | 0.11 | 2.17 | 3.39 |
| 1.86 | 3.16908 | 0.34076 | 1.56 | 3.33 | 2.94 | 0.10 | 2.18 | 3.40 |
| 1.87 | 3.16739 | 0.33714 | 1.57 | 3.35 | 2.93 | 0.10 | 2.19 | 3.42 |
| 1.88 | 3.16572 | 0.3336 | 1.58 | 3.37 | 2.93 | 0.10 | 2.20 | 3.43 |
| 1.89 | 3.16408 | 0.33013 | 1.59 | 3.39 | 2.93 | 0.10 | 2.22 | 3.45 |
| 1.9 | 3.16246 | 0.32673 | 1.60 | 3.41 | 2.93 | 0.10 | 2.23 | 3.46 |
| 1.91 | 3.16087 | 0.3234 | 1.60 | 3.43 | 2.93 | 0.10 | 2.24 | 3.48 |
| 1.92 | 3.15929 | 0.32014 | 1.61 | 3.45 | 2.93 | 0.09 | 2.25 | 3.50 |
| 1.93 | 3.15774 | 0.31695 | 1.62 | 3.47 | 2.93 | 0.09 | 2.26 | 3.51 |
| 1.94 | 3.15621 | 0.31382 | 1.63 | 3.49 | 2.92 | 0.09 | 2.27 | 3.53 |
| 1.95 | 3.15471 | 0.31075 | 1.64 | 3.51 | 2.92 | 0.09 | 2.28 | 3.54 |
| 1.96 | 3.15323 | 0.30775 | 1.65 | 3.52 | 2.92 | 0.09 | 2.29 | 3.56 |
| 1.97 | 3.15177 | 0.3048 | 1.66 | 3.54 | 2.92 | 0.09 | 2.30 | 3.57 |
| 1.98 | 3.15033 | 0.30191 | 1.67 | 3.56 | 2.92 | 0.08 | 2.32 | 3.59 |
| 1.99 | 3.14892 | 0.29908 | 1.67 | 3.58 | 2.92 | 0.08 | 2.33 | 3.60 |
| 2 | 3.14753 | 0.2963 | 1.68 | 3.60 | 2.92 | 0.08 | 2.34 | 3.61 |
| 2.1 | 3.12215 | 0.27874 | 1.81 | 3.80 | 2.91 | 0.09 | 2.71 | 3.76 |
| 2.2 | 3.11062 | 0.25741 | 1.89 | 3.97 | 2.89 | 0.09 | 2.88 | 3.87 |
| 2.3 | 3.09977 | 0.23928 | 1.97 | 4.13 | 2.88 | 0.08 | 3.04 | 3.97 |
| 2.4 | 3.08955 | 0.22371 | 2.06 | 4.28 | 2.87 | 0.07 | 3.20 | 4.06 |
| 2.5 | 3.07987 | 0.21019 | 2.14 | 4.43 | 2.86 | 0.07 | 3.34 | 4.14 |
| 2.6 | 3.07067 | 0.19834 | 2.23 | 4.58 | 2.85 | 0.06 | 3.49 | 4.22 |
| 2.7 | 3.06186 | 0.18788 | 2.31 | 4.72 | 2.84 | 0.06 | 3.63 | 4.28 |
| 2.8 | 3.05339 | 0.17859 | 2.40 | 4.85 | 2.84 | 0.06 | 3.76 | 4.34 |
| 2.9 | 3.04519 | 0.17027 | 2.49 | 4.99 | 2.83 | 0.05 | 3.89 | 4.40 |
| 3 | 3.03722 | 0.16278 | 2.57 | 5.12 | 2.82 | 0.05 | 4.01 | 4.45 |
| 3.1 | 3.02942 | 0.15602 | 2.66 | 5.25 | 2.82 | 0.05 | 4.13 | 4.50 |
| 3.2 | 3.02176 | 0.14987 | 2.75 | 5.37 | 2.81 | 0.05 | 4.25 | 4.54 |
| 3.3 | 3.01421 | 0.14426 | 2.83 | 5.49 | 2.81 | 0.05 | 4.36 | 4.58 |
| 3.4 | 3.00673 | 0.13912 | 2.92 | 5.61 | 2.80 | 0.05 | 4.47 | 4.62 |
| 3.5 | 2.99929 | 0.1344 | 3.00 | 5.73 | 2.80 | 0.04 | 4.57 | 4.66 |
| 3.6 | 2.99188 | 0.13005 | 3.09 | 5.84 | 2.79 | 0.04 | 4.68 | 4.69 |
| 3.7 | 2.98448 | 0.12603 | 3.17 | 5.95 | 2.79 | 0.04 | 4.78 | 4.72 |
| 3.8 | 2.97705 | 0.12231 | 3.26 | 6.06 | 2.79 | 0.04 | 4.87 | 4.75 |
| 3.9 | 2.96959 | 0.11886 | 3.34 | 6.17 | 2.78 | 0.04 | 4.97 | 4.78 |
| 4 | 2.96209 | 0.11564 | 3.42 | 6.27 | 2.78 | 0.04 | 5.06 | 4.81 |
| 4.1 | 2.95452 | 0.11265 | 3.51 | 6.37 | 2.77 | 0.04 | 5.15 | 4.83 |
| 4.2 | 2.94689 | 0.10985 | 3.59 | 6.47 | 2.77 | 0.04 | 5.23 | 4.86 |
| 4.3 | 2.93916 | 0.10724 | 3.67 | 6.57 | 2.77 | 0.04 | 5.32 | 4.88 |
| 4.4 | 2.93135 | 0.1048 | 3.75 | 6.67 | 2.76 | 0.04 | 5.40 | 4.90 |
| 4.5 | 2.92343 | 0.10251 | 3.83 | 6.76 | 2.76 | 0.04 | 5.48 | 4.92 |
| 4.6 | 2.9154 | 0.10036 | 3.91 | 6.86 | 2.75 | 0.04 | 5.56 | 4.94 |
| 4.7 | 2.90726 | 0.09835 | 3.99 | 6.95 | 2.75 | 0.04 | 5.64 | 4.96 |
| 4.8 | 2.89898 | 0.09646 | 4.07 | 7.04 | 2.74 | 0.04 | 5.71 | 4.98 |
| 4.9 | 2.89057 | 0.09468 | 4.14 | 7.13 | 2.74 | 0.04 | 5.79 | 5.00 |
| 5 | 2.88203 | 0.09302 | 4.22 | 7.22 | 2.73 | 0.04 | 5.86 | 5.02 |
| 5.1 | 2.87333 | 0.09145 | 4.30 | 7.30 | 2.73 | 0.03 | 5.93 | 5.04 |
| 5.2 | 2.86449 | 0.08998 | 4.37 | 7.39 | 2.72 | 0.03 | 6.00 | 5.05 |
| 5.3 | 2.85548 | 0.0886 | 4.45 | 7.47 | 2.72 | 0.03 | 6.07 | 5.07 |
| 5.4 | 2.84632 | 0.08731 | 4.52 | 7.56 | 2.71 | 0.03 | 6.14 | 5.08 |
| 5.5 | 2.83698 | 0.08609 | 4.59 | 7.64 | 2.71 | 0.03 | 6.21 | 5.10 |
| 5.6 | 2.82747 | 0.08495 | 4.67 | 7.72 | 2.70 | 0.03 | 6.28 | 5.11 |
| 5.7 | 2.81778 | 0.08389 | 4.74 | 7.80 | 2.70 | 0.03 | 6.35 | 5.12 |
| 5.8 | 2.80791 | 0.0829 | 4.81 | 7.88 | 2.69 | 0.03 | 6.41 | 5.13 |
| 5.9 | 2.79784 | 0.08197 | 4.88 | 7.96 | 2.69 | 0.03 | 6.48 | 5.14 |
| 6 | 2.78758 | 0.08111 | 4.95 | 8.03 | 2.68 | 0.03 | 6.54 | 5.15 |
| 6.1 | 2.77712 | 0.08032 | 5.02 | 8.11 | 2.68 | 0.03 | 6.61 | 5.16 |
| 6.2 | 2.76646 | 0.07958 | 5.09 | 8.19 | 2.67 | 0.03 | 6.67 | 5.17 |
| 6.3 | 2.75559 | 0.0789 | 5.15 | 8.26 | 2.66 | 0.03 | 6.73 | 5.17 |
| 6.4 | 2.74449 | 0.07828 | 5.22 | 8.34 | 2.66 | 0.03 | 6.79 | 5.17 |
| 6.5 | 2.73318 | 0.07772 | 5.29 | 8.41 | 2.65 | 0.03 | 6.85 | 5.17 |
| 6.6 | 2.72164 | 0.07721 | 5.35 | 8.48 | 2.64 | 0.03 | 6.91 | 5.17 |
| 6.7 | 2.70987 | 0.07676 | 5.42 | 8.56 | 2.64 | 0.03 | 6.97 | 5.17 |
| 6.8 | 2.69786 | 0.07636 | 5.48 | 8.63 | 2.63 | 0.03 | 7.02 | 5.17 |
| 6.9 | 2.68561 | 0.07601 | 5.55 | 8.70 | 2.62 | 0.03 | 7.08 | 5.17 |
| 7 | 2.6731 | 0.07571 | 5.61 | 8.77 | 2.62 | 0.03 | 7.13 | 5.16 |
| 7.1 | 2.66034 | 0.07547 | 5.68 | 8.84 | 2.61 | 0.03 | 7.18 | 5.16 |
| 7.2 | 2.64732 | 0.07528 | 5.74 | 8.91 | 2.60 | 0.03 | 7.23 | 5.15 |
| 7.3 | 2.63402 | 0.07514 | 5.81 | 8.98 | 2.59 | 0.03 | 7.27 | 5.15 |
| 7.4 | 2.62045 | 0.07505 | 5.87 | 9.05 | 2.59 | 0.03 | 7.32 | 5.14 |
| 7.5 | 2.60659 | 0.07501 | 5.93 | 9.12 | 2.58 | 0.03 | 7.36 | 5.13 |
| 7.6 | 2.59244 | 0.07503 | 5.99 | 9.19 | 2.57 | 0.03 | 7.41 | 5.12 |
| 7.7 | 2.57799 | 0.07509 | 6.06 | 9.26 | 2.56 | 0.03 | 7.45 | 5.11 |
| 7.8 | 2.56323 | 0.07522 | 6.12 | 9.33 | 2.55 | 0.03 | 7.49 | 5.10 |
| 7.9 | 2.54815 | 0.07539 | 6.18 | 9.40 | 2.54 | 0.03 | 7.53 | 5.09 |
| 8 | 2.53275 | 0.07562 | 6.24 | 9.47 | 2.53 | 0.03 | 7.56 | 5.08 |
| 8.1 | 2.51701 | 0.07591 | 6.30 | 9.53 | 2.52 | 0.03 | 7.60 | 5.07 |
| 8.2 | 2.50092 | 0.07626 | 6.37 | 9.60 | 2.51 | 0.03 | 7.63 | 5.06 |
| 8.3 | 2.48448 | 0.07667 | 6.43 | 9.67 | 2.50 | 0.03 | 7.67 | 5.05 |
| 8.4 | 2.46766 | 0.07714 | 6.49 | 9.73 | 2.49 | 0.03 | 7.70 | 5.03 |
| 8.5 | 2.45047 | 0.07767 | 6.55 | 9.80 | 2.48 | 0.03 | 7.74 | 5.02 |
| 8.6 | 2.43288 | 0.07827 | 6.61 | 9.86 | 2.47 | 0.03 | 7.77 | 5.00 |
| 8.7 | 2.41489 | 0.07894 | 6.68 | 9.93 | 2.46 | 0.03 | 7.80 | 4.98 |
| 8.8 | 2.39647 | 0.07968 | 6.74 | 9.99 | 2.44 | 0.03 | 7.84 | 4.96 |
| 8.9 | 2.37762 | 0.0805 | 6.80 | 10.06 | 2.43 | 0.03 | 7.89 | 4.93 |
| 9 | 2.35832 | 0.08139 | 6.86 | 10.12 | 2.42 | 0.03 | 7.93 | 4.88 |
| 9.1 | 2.33855 | 0.08237 | 6.92 | 10.18 | 2.40 | 0.03 | 7.98 | 4.81 |
| 9.2 | 2.31829 | 0.08343 | 6.99 | 10.24 | 2.39 | 0.03 | 8.01 | 4.70 |
| 9.3 | 2.29752 | 0.08459 | 7.05 | 10.30 | 2.38 | 0.03 | 7.99 | 4.56 |
| 9.4 | 2.27623 | 0.08585 | 7.11 | 10.36 | 2.36 | 0.03 | 7.90 | 4.42 |
| 9.5 | 2.25439 | 0.08721 | 7.17 | 10.42 | 2.34 | 0.03 | 7.73 | 4.31 |
| 9.6 | 2.23197 | 0.08868 | 7.23 | 10.48 | 2.33 | 0.03 | 7.52 | 4.28 |
| 9.7 | 2.20896 | 0.09027 | 7.29 | 10.54 | 2.31 | 0.03 | 7.29 | 4.34 |
| 9.8 | 2.18532 | 0.09198 | 7.36 | 10.60 | 2.29 | 0.03 | 7.09 | 4.48 |
| 9.9 | 2.16102 | 0.09383 | 7.42 | 10.65 | 2.27 | 0.03 | 6.96 | 4.71 |
| 10 | 2.13604 | 0.09583 | 7.48 | 10.71 | 2.25 | 0.03 | 6.91 | 4.98 |
| 10.1 | 2.11033 | 0.09799 | 7.54 | 10.77 | 2.23 | 0.03 | 6.99 | 5.24 |
| 10.2 | 2.08386 | 0.10032 | 7.60 | 10.82 | 2.21 | 0.04 | 7.17 | 5.44 |
| 10.3 | 2.0566 | 0.10284 | 7.66 | 10.87 | 2.19 | 0.04 | 7.40 | 5.51 |
| 10.4 | 2.02848 | 0.10556 | 7.72 | 10.93 | 2.16 | 0.04 | 7.60 | 5.47 |
| 10.5 | 1.99947 | 0.10851 | 7.78 | 10.98 | 2.14 | 0.04 | 7.71 | 5.36 |
| 10.6 | 1.9695 | 0.1117 | 7.84 | 11.03 | 2.11 | 0.04 | 7.73 | 5.23 |
| 10.7 | 1.93853 | 0.11517 | 7.90 | 11.08 | 2.08 | 0.04 | 7.67 | 5.14 |
| 10.8 | 1.90649 | 0.11894 | 7.95 | 11.13 | 2.05 | 0.05 | 7.58 | 5.12 |
| 10.9 | 1.87329 | 0.12305 | 8.01 | 11.18 | 2.02 | 0.05 | 7.50 | 5.15 |
| 11 | 1.83887 | 0.12753 | 8.07 | 11.23 | 1.98 | 0.06 | 7.43 | 5.21 |
| 11.1 | 1.80314 | 0.13244 | 8.13 | 11.27 | 1.94 | 0.06 | 7.40 | 5.29 |
| 11.2 | 1.76599 | 0.13783 | 8.18 | 11.32 | 1.90 | 0.07 | 7.38 | 5.38 |
| 11.3 | 1.72731 | 0.14377 | 8.24 | 11.37 | 1.86 | 0.08 | 7.38 | 5.46 |
| 11.4 | 1.68698 | 0.15033 | 8.29 | 11.42 | 1.81 | 0.10 | 7.39 | 5.53 |
| 11.5 | 1.64485 | 0.15762 | 8.35 | 11.46 | 1.76 | 0.12 | 7.41 | 5.60 |
| 11.6 | 1.60077 | 0.16574 | 8.40 | 11.51 | 1.71 | 0.14 | 7.43 | 5.67 |
| 11.7 | 1.55454 | 0.17485 | 8.45 | 11.55 | 1.66 | 0.17 | 7.45 | 5.73 |
| 11.8 | 1.50597 | 0.18513 | 8.50 | 11.60 | 1.60 | 0.20 | 7.48 | 5.79 |
| 11.9 | 1.45481 | 0.1968 | 8.56 | 11.64 | 1.54 | 0.24 | 7.51 | 5.85 |
| 12 | 1.4008 | 0.21016 | 8.61 | 11.69 | 1.49 | 0.29 | 7.54 | 5.91 |
| 12.1 | 1.34362 | 0.22558 | 8.66 | 11.73 | 1.43 | 0.35 | 7.57 | 5.96 |
| 12.2 | 1.28295 | 0.24358 | 8.71 | 11.78 | 1.38 | 0.41 | 7.61 | 6.01 |
| 12.3 | 1.21842 | 0.26482 | 8.75 | 11.82 | 1.33 | 0.49 | 7.64 | 6.06 |
| 12.4 | 1.14968 | 0.29024 | 8.80 | 11.86 | 1.30 | 0.57 | 7.68 | 6.11 |
| 12.5 | 1.07647 | 0.32109 | 8.85 | 11.91 | 1.27 | 0.66 | 7.72 | 6.16 |
| 12.6 | 0.99876 | 0.35909 | 8.90 | 11.95 | 1.25 | 0.76 | 7.76 | 6.20 |
| 12.7 | 0.91712 | 0.40653 | 8.95 | 11.99 | 1.25 | 0.85 | 7.79 | 6.25 |
| 12.8 | 0.83326 | 0.46606 | 8.99 | 12.04 | 1.25 | 0.94 | 7.83 | 6.29 |
| 12.9 | 0.75069 | 0.53997 | 9.04 | 12.08 | 1.26 | 1.03 | 7.87 | 6.33 |
| 13 | 0.67459 | 0.6286 | 9.08 | 12.12 | 1.29 | 1.11 | 7.91 | 6.37 |
| 13.1 | 0.60994 | 0.72905 | 9.13 | 12.17 | 1.31 | 1.19 | 7.95 | 6.41 |
| 13.2 | 0.55903 | 0.8363 | 9.17 | 12.21 | 1.34 | 1.26 | 7.99 | 6.44 |
| 13.3 | 0.52115 | 0.94576 | 9.22 | 12.26 | 1.38 | 1.33 | 8.03 | 6.48 |
| 13.4 | 0.49417 | 1.05467 | 9.26 | 12.30 | 1.42 | 1.39 | 8.07 | 6.51 |
| 13.5 | 0.47587 | 1.16185 | 9.31 | 12.34 | 1.46 | 1.45 | 8.11 | 6.55 |
| 13.6 | 0.46452 | 1.26705 | 9.35 | 12.39 | 1.50 | 1.50 | 8.15 | 6.58 |
| 13.7 | 0.45889 | 1.37049 | 9.39 | 12.43 | 1.54 | 1.55 | 8.20 | 6.61 |
| 13.8 | 0.45822 | 1.47256 | 9.43 | 12.48 | 1.58 | 1.59 | 8.24 | 6.64 |
| 13.9 | 0.46206 | 1.57373 | 9.48 | 12.52 | 1.62 | 1.63 | 8.28 | 6.67 |
| 14 | 0.47024 | 1.67447 | 9.52 | 12.56 | 1.66 | 1.66 | 8.32 | 6.70 |
| 14.1 | 0.48279 | 1.77522 | 9.56 | 12.61 | 1.70 | 1.69 | 8.36 | 6.73 |
| 14.2 | 0.49996 | 1.8764 | 9.60 | 12.65 | 1.73 | 1.73 | 8.40 | 6.75 |
| 14.3 | 0.52218 | 1.97837 | 9.65 | 12.70 | 1.77 | 1.75 | 8.44 | 6.78 |
| 14.4 | 0.55012 | 2.08142 | 9.69 | 12.74 | 1.80 | 1.78 | 8.48 | 6.81 |
| 14.5 | 0.58466 | 2.18571 | 9.73 | 12.79 | 1.82 | 1.82 | 8.52 | 6.83 |
| 14.6 | 0.62694 | 2.29122 | 9.77 | 12.83 | 1.84 | 1.85 | 8.56 | 6.86 |
| 14.7 | 0.67842 | 2.39765 | 9.81 | 12.88 | 1.87 | 1.89 | 8.60 | 6.88 |
| 14.8 | 0.74083 | 2.50426 | 9.85 | 12.92 | 1.89 | 1.94 | 8.64 | 6.90 |
| 14.9 | 0.81618 | 2.60962 | 9.89 | 12.97 | 1.92 | 1.99 | 8.68 | 6.92 |
| 15 | 0.90654 | 2.71127 | 9.93 | 13.01 | 1.95 | 2.06 | 8.72 | 6.95 |
| 15.1 | 1.01362 | 2.80544 | 9.98 | 13.06 | 2.00 | 2.12 | 8.76 | 6.97 |
| 15.2 | 1.13795 | 2.88677 | 10.02 | 13.10 | 2.06 | 2.19 | 8.79 | 6.99 |
| 15.3 | 1.27756 | 2.94861 | 10.06 | 13.15 | 2.13 | 2.26 | 8.83 | 7.01 |
| 15.4 | 1.42662 | 2.98453 | 10.10 | 13.19 | 2.22 | 2.32 | 8.87 | 7.03 |
| 15.5 | 1.57504 | 2.99127 | 10.14 | 13.24 | 2.33 | 2.37 | 8.91 | 7.04 |
| 15.6 | 1.71084 | 2.97232 | 10.18 | 13.28 | 2.44 | 2.40 | 8.95 | 7.06 |
| 15.7 | 1.82584 | 2.93932 | 10.22 | 13.33 | 2.57 | 2.42 | 8.98 | 7.08 |
| 15.8 | 1.92195 | 2.90802 | 10.26 | 13.37 | 2.70 | 2.42 | 9.02 | 7.10 |
| 15.9 | 2.01316 | 2.88978 | 10.30 | 13.42 | 2.83 | 2.39 | 9.06 | 7.11 |
| 16 | 2.1206 | 2.88269 | 10.34 | 13.46 | 2.95 | 2.35 | 9.09 | 7.13 |
| 16.1 | 2.26165 | 2.86756 | 10.38 | 13.51 | 3.07 | 2.29 | 9.13 | 7.15 |
| 16.2 | 2.43661 | 2.81217 | 10.42 | 13.55 | 3.17 | 2.21 | 9.17 | 7.16 |
| 16.3 | 2.6195 | 2.68689 | 10.46 | 13.60 | 3.26 | 2.11 | 9.20 | 7.18 |
| 16.4 | 2.76522 | 2.48699 | 10.51 | 13.64 | 3.33 | 2.01 | 9.24 | 7.19 |
| 16.5 | 2.83611 | 2.24099 | 10.55 | 13.68 | 3.38 | 1.90 | 9.27 | 7.21 |
| 16.6 | 2.82377 | 1.99176 | 10.59 | 13.73 | 3.42 | 1.78 | 9.31 | 7.22 |
| 16.7 | 2.7446 | 1.77135 | 10.63 | 13.77 | 3.43 | 1.66 | 9.34 | 7.24 |
| 16.8 | 2.62216 | 1.59323 | 10.67 | 13.82 | 3.43 | 1.55 | 9.38 | 7.25 |
| 16.9 | 2.47571 | 1.45821 | 10.71 | 13.86 | 3.41 | 1.44 | 9.41 | 7.26 |
| 17 | 2.31772 | 1.36194 | 10.75 | 13.91 | 3.37 | 1.34 | 9.45 | 7.28 |
| 17.1 | 2.15544 | 1.29925 | 10.79 | 13.95 | 3.32 | 1.25 | 9.48 | 7.29 |
| 17.2 | 1.99294 | 1.26597 | 10.83 | 13.99 | 3.26 | 1.17 | 9.51 | 7.30 |
| 17.3 | 1.83284 | 1.25927 | 10.87 | 14.04 | 3.18 | 1.11 | 9.55 | 7.31 |
| 17.4 | 1.67745 | 1.2775 | 10.91 | 14.08 | 3.10 | 1.05 | 9.58 | 7.32 |
| 17.5 | 1.52946 | 1.31957 | 10.95 | 14.13 | 3.01 | 1.01 | 9.61 | 7.34 |
| 17.6 | 1.39226 | 1.38417 | 11.00 | 14.17 | 2.92 | 0.99 | 9.64 | 7.35 |
| 17.7 | 1.26953 | 1.46901 | 11.04 | 14.21 | 2.82 | 0.99 | 9.67 | 7.36 |
| 17.8 | 1.16444 | 1.57056 | 11.08 | 14.25 | 2.72 | 1.00 | 9.71 | 7.37 |
| 17.9 | 1.07881 | 1.68456 | 11.12 | 14.30 | 2.63 | 1.03 | 9.74 | 7.38 |
| 18 | 1.01301 | 1.80693 | 11.16 | 14.34 | 2.54 | 1.08 | 9.77 | 7.39 |
| 18.1 | 0.96648 | 1.93448 | 11.20 | 14.38 | 2.47 | 1.15 | 9.80 | 7.40 |
| 18.2 | 0.93847 | 2.06507 | 11.24 | 14.43 | 2.41 | 1.24 | 9.83 | 7.41 |
| 18.3 | 0.92865 | 2.19731 | 11.28 | 14.47 | 2.37 | 1.33 | 9.86 | 7.42 |
| 18.4 | 0.93756 | 2.33014 | 11.32 | 14.51 | 2.36 | 1.43 | 9.89 | 7.43 |
| 18.5 | 0.96677 | 2.46216 | 11.36 | 14.55 | 2.37 | 1.53 | 9.92 | 7.44 |
| 18.6 | 1.01893 | 2.59074 | 11.40 | 14.59 | 2.39 | 1.63 | 9.95 | 7.45 |
| 18.7 | 1.0972 | 2.71073 | 11.44 | 14.64 | 2.44 | 1.71 | 9.98 | 7.46 |
| 18.8 | 1.20333 | 2.81318 | 11.48 | 14.68 | 2.50 | 1.78 | 10.01 | 7.47 |
| 18.9 | 1.33366 | 2.88554 | 11.52 | 14.72 | 2.58 | 1.84 | 10.04 | 7.47 |
| 19 | 1.47482 | 2.91703 | 11.56 | 14.76 | 2.66 | 1.88 | 10.07 | 7.48 |
| 19.1 | 1.60726 | 2.90895 | 11.61 | 14.80 | 2.74 | 1.90 | 10.09 | 7.49 |
| 19.2 | 1.7202 | 2.87622 | 11.65 | 14.84 | 2.82 | 1.91 | 10.12 | 7.50 |
| 19.3 | 1.81955 | 2.83118 | 11.69 | 14.88 | 2.90 | 1.90 | 10.15 | 7.51 |
| 19.4 | 1.91445 | 2.77048 | 11.73 | 14.93 | 2.98 | 1.88 | 10.18 | 7.52 |
| 19.5 | 2.0015 | 2.6825 | 11.77 | 14.97 | 3.05 | 1.84 | 10.20 | 7.52 |
| 19.6 | 2.06447 | 2.56278 | 11.81 | 15.01 | 3.11 | 1.79 | 10.23 | 7.53 |
| 19.7 | 2.08587 | 2.42123 | 11.85 | 15.05 | 3.16 | 1.74 | 10.26 | 7.54 |
| 19.8 | 2.05792 | 2.27749 | 11.89 | 15.09 | 3.20 | 1.68 | 10.29 | 7.55 |
| 19.9 | 1.98503 | 2.15112 | 11.93 | 15.13 | 3.23 | 1.62 | 10.31 | 7.55 |
| 20 | 1.87918 | 2.05517 | 11.97 | 15.17 | 3.24 | 1.55 | 10.34 | 7.56 |
| 20.1 | 1.75432 | 1.99529 | 12.01 | 15.21 | 3.25 | 1.48 | 10.36 | 7.57 |
| 20.2 | 1.62309 | 1.97159 | 12.05 | 15.25 | 3.24 | 1.42 | 10.39 | 7.58 |
| 20.3 | 1.49578 | 1.98077 | 12.09 | 15.28 | 3.22 | 1.35 | 10.42 | 7.58 |
| 20.4 | 1.3801 | 2.01742 | 12.13 | 15.32 | 3.20 | 1.30 | 10.44 | 7.59 |
| 20.5 | 1.28129 | 2.07504 | 12.17 | 15.36 | 3.16 | 1.24 | 10.47 | 7.60 |
| 20.6 | 1.20225 | 2.1468 | 12.21 | 15.40 | 3.12 | 1.20 | 10.49 | 7.60 |
| 20.7 | 1.14406 | 2.22607 | 12.25 | 15.44 | 3.07 | 1.16 | 10.52 | 7.61 |
| 20.8 | 1.10634 | 2.3065 | 12.29 | 15.48 | 3.02 | 1.13 | 10.54 | 7.62 |
| 20.9 | 1.08727 | 2.38181 | 12.33 | 15.52 | 2.96 | 1.11 | 10.56 | 7.63 |
| 21 | 1.08293 | 2.44586 | 12.37 | 15.56 | 2.91 | 1.10 | 10.59 | 7.63 |
| 21.1 | 1.08633 | 2.49384 | 12.41 | 15.59 | 2.85 | 1.10 | 10.61 | 7.64 |
| 21.2 | 1.08771 | 2.52489 | 12.45 | 15.63 | 2.79 | 1.10 | 10.64 | 7.64 |
| 21.3 | 1.07785 | 2.54439 | 12.48 | 15.67 | 2.74 | 1.12 | 10.66 | 7.65 |
| 21.4 | 1.0529 | 2.56239 | 12.52 | 15.71 | 2.69 | 1.14 | 10.68 | 7.66 |
| 21.5 | 1.01611 | 2.58823 | 12.56 | 15.75 | 2.65 | 1.16 | 10.71 | 7.66 |
| 21.6 | 0.97457 | 2.62634 | 12.60 | 15.78 | 2.61 | 1.19 | 10.73 | 7.67 |
| 21.7 | 0.93487 | 2.67623 | 12.64 | 15.82 | 2.58 | 1.23 | 10.75 | 7.68 |
| 21.8 | 0.90102 | 2.73498 | 12.68 | 15.86 | 2.55 | 1.26 | 10.77 | 7.68 |
| 21.9 | 0.87463 | 2.79932 | 12.72 | 15.90 | 2.53 | 1.30 | 10.80 | 7.69 |
| 22 | 0.85587 | 2.86663 | 12.76 | 15.93 | 2.52 | 1.33 | 10.82 | 7.69 |
| 22.1 | 0.84425 | 2.93507 | 12.80 | 15.97 | 2.51 | 1.37 | 10.84 | 7.70 |
| 22.2 | 0.83907 | 3.00344 | 12.84 | 16.01 | 2.50 | 1.40 | 10.86 | 7.71 |
| 22.3 | 0.83969 | 3.07096 | 12.87 | 16.04 | 2.50 | 1.43 | 10.88 | 7.71 |
| 22.4 | 0.84556 | 3.13711 | 12.91 | 16.08 | 2.50 | 1.46 | 10.91 | 7.72 |
| 22.5 | 0.85623 | 3.20152 | 12.95 | 16.11 | 2.51 | 1.49 | 10.93 | 7.72 |
| 22.6 | 0.87137 | 3.2639 | 12.99 | 16.15 | 2.51 | 1.51 | 10.95 | 7.73 |
| 22.7 | 0.89073 | 3.32395 | 13.03 | 16.19 | 2.52 | 1.53 | 10.97 | 7.74 |
| 22.8 | 0.91411 | 3.38135 | 13.07 | 16.22 | 2.53 | 1.55 | 10.99 | 7.74 |
| 22.9 | 0.94132 | 3.43575 | 13.10 | 16.26 | 2.53 | 1.56 | 11.01 | 7.75 |
| 23 | 0.97217 | 3.48674 | 13.14 | 16.29 | 2.54 | 1.57 | 11.03 | 7.75 |
| 23.1 | 1.00642 | 3.53385 | 13.18 | 16.33 | 2.55 | 1.58 | 11.05 | 7.76 |
| 23.2 | 1.04376 | 3.57656 | 13.22 | 16.37 | 2.56 | 1.59 | 11.07 | 7.77 |
| 23.3 | 1.08376 | 3.6143 | 13.26 | 16.40 | 2.56 | 1.59 | 11.09 | 7.77 |
| 23.4 | 1.12586 | 3.6465 | 13.29 | 16.44 | 2.56 | 1.59 | 11.11 | 7.78 |
| 23.5 | 1.16933 | 3.67262 | 13.33 | 16.47 | 2.57 | 1.59 | 11.13 | 7.78 |
| 23.6 | 1.21325 | 3.69218 | 13.37 | 16.51 | 2.57 | 1.59 | 11.15 | 7.79 |
| 23.7 | 1.2565 | 3.7049 | 13.41 | 16.54 | 2.56 | 1.59 | 11.17 | 7.79 |
| 23.8 | 1.29783 | 3.71067 | 13.44 | 16.58 | 2.56 | 1.59 | 11.19 | 7.80 |
| 23.9 | 1.33588 | 3.70971 | 13.48 | 16.61 | 2.55 | 1.58 | 11.21 | 7.81 |
| 24 | 1.36928 | 3.70254 | 13.52 | 16.64 | 2.54 | 1.58 | 11.23 | 7.81 |
| 24.1 | 1.39677 | 3.69008 | 13.56 | 16.68 | 2.53 | 1.57 | 11.25 | 7.82 |
| 24.2 | 1.41726 | 3.67354 | 13.59 | 16.71 | 2.51 | 1.56 | 11.26 | 7.82 |
| 24.3 | 1.43001 | 3.65442 | 13.63 | 16.75 | 2.50 | 1.56 | 11.28 | 7.83 |
| 24.4 | 1.43465 | 3.63435 | 13.67 | 16.78 | 2.48 | 1.56 | 11.30 | 7.83 |
| 24.5 | 1.43123 | 3.61502 | 13.71 | 16.82 | 2.45 | 1.55 | 11.32 | 7.84 |
| 24.6 | 1.42021 | 3.598 | 13.74 | 16.85 | 2.42 | 1.55 | 11.34 | 7.85 |
| 24.7 | 1.40238 | 3.58469 | 13.78 | 16.88 | 2.39 | 1.55 | 11.36 | 7.85 |
| 24.8 | 1.37879 | 3.57617 | 13.82 | 16.92 | 2.36 | 1.55 | 11.37 | 7.86 |
| 24.9 | 1.35065 | 3.57323 | 13.85 | 16.95 | 2.33 | 1.55 | 11.39 | 7.86 |
| 25 | 1.31922 | 3.57633 | 13.89 | 16.98 | 2.29 | 1.56 | 11.41 | 7.87 |
| 25.1 | 1.28572 | 3.58561 | 13.93 | 17.02 | 2.25 | 1.57 | 11.43 | 7.87 |
| 25.2 | 1.25129 | 3.60099 | 13.96 | 17.05 | 2.21 | 1.58 | 11.45 | 7.88 |
| 25.3 | 1.2169 | 3.62216 | 14.00 | 17.08 | 2.16 | 1.60 | 11.46 | 7.88 |
| 25.4 | 1.18338 | 3.64871 | 14.03 | 17.12 | 2.12 | 1.62 | 11.48 | 7.89 |
| 25.5 | 1.15136 | 3.68011 | 14.07 | 17.15 | 2.07 | 1.65 | 11.50 | 7.90 |
| 25.6 | 1.12135 | 3.71583 | 14.11 | 17.18 | 2.02 | 1.68 | 11.51 | 7.90 |
| 25.7 | 1.09367 | 3.75533 | 14.14 | 17.22 | 1.97 | 1.71 | 11.53 | 7.91 |
| 25.8 | 1.06857 | 3.79808 | 14.18 | 17.25 | 1.92 | 1.76 | 11.55 | 7.91 |
| 25.9 | 1.04616 | 3.84361 | 14.21 | 17.28 | 1.88 | 1.80 | 11.56 | 7.92 |
| 26 | 1.02651 | 3.89149 | 14.25 | 17.31 | 1.83 | 1.85 | 11.58 | 7.92 |
| 26.1 | 1.00962 | 3.94135 | 14.29 | 17.35 | 1.79 | 1.91 | 11.60 | 7.93 |
| 26.2 | 0.99547 | 3.99288 | 14.32 | 17.38 | 1.75 | 1.97 | 11.61 | 7.93 |
| 26.3 | 0.98398 | 4.04579 | 14.36 | 17.41 | 1.71 | 2.04 | 11.63 | 7.94 |
| 26.4 | 0.9751 | 4.09986 | 14.39 | 17.44 | 1.68 | 2.11 | 11.65 | 7.95 |
| 26.5 | 0.96875 | 4.15492 | 14.43 | 17.47 | 1.65 | 2.18 | 11.66 | 7.95 |
| 26.6 | 0.96486 | 4.21079 | 14.46 | 17.51 | 1.63 | 2.26 | 11.68 | 7.96 |
| 26.7 | 0.96334 | 4.26737 | 14.50 | 17.54 | 1.61 | 2.33 | 11.69 | 7.96 |
| 26.8 | 0.96415 | 4.32455 | 14.53 | 17.57 | 1.60 | 2.41 | 11.71 | 7.97 |
| 26.9 | 0.96723 | 4.38225 | 14.57 | 17.60 | 1.59 | 2.50 | 11.73 | 7.97 |
| 27 | 0.97254 | 4.44042 | 14.60 | 17.63 | 1.59 | 2.58 | 11.74 | 7.98 |
| 27.1 | 0.98004 | 4.49899 | 14.64 | 17.67 | 1.59 | 2.66 | 11.76 | 7.99 |
| 27.2 | 0.98973 | 4.55793 | 14.67 | 17.70 | 1.59 | 2.74 | 11.77 | 7.99 |
| 27.3 | 1.00159 | 4.61721 | 14.71 | 17.73 | 1.60 | 2.82 | 11.79 | 8.00 |
| 27.4 | 1.01564 | 4.67679 | 14.74 | 17.76 | 1.61 | 2.90 | 11.80 | 8.00 |
| 27.5 | 1.03188 | 4.73666 | 14.78 | 17.79 | 1.63 | 2.98 | 11.82 | 8.01 |
| 27.6 | 1.05036 | 4.7968 | 14.81 | 17.82 | 1.65 | 3.06 | 11.83 | 8.01 |
| 27.7 | 1.07113 | 4.85717 | 14.85 | 17.85 | 1.68 | 3.14 | 11.85 | 8.02 |
| 27.8 | 1.09423 | 4.91776 | 14.88 | 17.88 | 1.70 | 3.21 | 11.86 | 8.03 |
| 27.9 | 1.11975 | 4.97853 | 14.91 | 17.91 | 1.73 | 3.29 | 11.88 | 8.03 |
| 28 | 1.14776 | 5.03945 | 14.95 | 17.95 | 1.77 | 3.36 | 11.89 | 8.04 |
| 28.1 | 1.17836 | 5.10047 | 14.98 | 17.98 | 1.80 | 3.43 | 11.91 | 8.04 |
| 28.2 | 1.21166 | 5.16155 | 15.02 | 18.01 | 1.84 | 3.50 | 11.92 | 8.05 |
| 28.3 | 1.24779 | 5.22261 | 15.05 | 18.04 | 1.88 | 3.57 | 11.93 | 8.06 |
| 28.4 | 1.28687 | 5.28359 | 15.09 | 18.07 | 1.93 | 3.64 | 11.95 | 8.06 |
| 28.5 | 1.32905 | 5.34439 | 15.12 | 18.10 | 1.97 | 3.70 | 11.96 | 8.07 |
| 28.6 | 1.3745 | 5.40489 | 15.15 | 18.13 | 2.02 | 3.76 | 11.98 | 8.07 |
| 28.7 | 1.42336 | 5.46496 | 15.19 | 18.16 | 2.07 | 3.82 | 11.99 | 8.08 |
| 28.8 | 1.47584 | 5.52444 | 15.22 | 18.19 | 2.12 | 3.88 | 12.01 | 8.09 |
| 28.9 | 1.5321 | 5.58314 | 15.25 | 18.22 | 2.17 | 3.94 | 12.02 | 8.09 |
| 29 | 1.59234 | 5.64083 | 15.29 | 18.25 | 2.23 | 3.99 | 12.03 | 8.10 |
| 29.1 | 1.65676 | 5.69726 | 15.32 | 18.28 | 2.28 | 4.05 | 12.05 | 8.10 |
| 29.2 | 1.72553 | 5.75212 | 15.35 | 18.31 | 2.34 | 4.10 | 12.06 | 8.11 |
| 29.3 | 1.79885 | 5.80507 | 15.39 | 18.34 | 2.40 | 4.15 | 12.07 | 8.12 |
| 29.4 | 1.87687 | 5.85569 | 15.42 | 18.37 | 2.46 | 4.19 | 12.09 | 8.12 |
| 29.5 | 1.95975 | 5.90355 | 15.45 | 18.40 | 2.52 | 4.24 | 12.10 | 8.13 |
| 29.6 | 2.04761 | 5.94812 | 15.49 | 18.43 | 2.58 | 4.28 | 12.11 | 8.13 |
| 29.7 | 2.1405 | 5.98884 | 15.52 | 18.46 | 2.64 | 4.32 | 12.13 | 8.14 |
| 29.8 | 2.23845 | 6.02507 | 15.55 | 18.49 | 2.71 | 4.36 | 12.14 | 8.15 |
| 29.9 | 2.34138 | 6.05611 | 15.59 | 18.52 | 2.77 | 4.40 | 12.15 | 8.15 |
| 30 | 2.44916 | 6.08124 | 15.62 | 18.55 | 2.84 | 4.43 | 12.17 | 8.16 |